\documentclass[preprint,review,authoryear,12pt,3p]{elsarticle}

\usepackage{amsmath}
\usepackage{graphicx,amssymb,psfrag,subfigure,color}
\usepackage{epsfig}
\newcommand{\beq}{\begin{equation}}
\newcommand{\eeq}{\end{equation}}
\newcommand{\bpm}{\begin{pmatrix}}
\newcommand{\epm}{\end{pmatrix}}
\newcommand{\beqa}{\begin{eqnarray}}
\newcommand{\eeqa}{\end{eqnarray}}
\newcommand{\beqas}{\begin{eqnarray*}}
\newcommand{\eeqas}{\end{eqnarray*}}

\newcommand{\im}{\mathrm{i}}
\renewcommand{\d}{\mathrm{d}}

\newcommand{\pdhfrac}[2]{\mathchoice{\frac{#1}{#2}}{#1/#2}{#1/#2}{#1/#2}}

\newcommand{\pd}[2]{\pdhfrac{{\partial}#1}{{\partial}#2}}
\newcommand{\spd}[2]{\pdhfrac{\partial^2#1}{{\partial}#2^2}}

\newcommand{\eps}{\epsilon}

\def\XXint#1#2#3{{\setbox0=\hbox{$#1{#2#3}{\int}$ }
\vcenter{\hbox{$#2#3$ }}\kern-.6\wd0}}

\journal{}

\begin{document}

\begin{frontmatter}

\title{Continuum dynamics of the formation, migration and dissociation of self-locked dislocation structures on parallel slip planes}
\author[HKUST]{Yichao~Zhu\corref{cor1}}
\ead{mayczhu@ust.hk}
\author[HKUST]{Xiaohua~Niu}
\author[HKUST]{Yang~Xiang}
%\ead{maxiang@ust.hk}

\cortext[cor1]{Corresponding author}
%\cortext[cor2]{Principal corresponding author}
%\fntext[fn1]{This is the specimen author footnote.}
%\fntext[fn2]{Another author footnote, but a little more longer.}
%\fntext[fn3]{Yet another author footnote. Indeed, you can have
%any number of author footnotes.}

\address[HKUST]{Department of Mathematics, The Hong Kong University of Science and Technology, Clear Water Bay, Kowloon, Hong Kong}

%\maketitle

\begin{abstract}
  In continuum models of dislocations, proper formulations of short-range elastic interactions of dislocations are crucial for capturing  various types of dislocation patterns formed in crystalline materials.
  In this article, the continuum dynamics of straight dislocations distributed on two parallel slip planes is modelled through upscaling the underlying discrete dislocation dynamics. Two continuum velocity field quantities are introduced to facilitate the discrete-to-continuum transition. The first one is the local migration velocity of dislocation ensembles which is found fully independent of the short-range dislocation correlations. The second one is the decoupling velocity of dislocation pairs controlled by a threshold stress value, which is proposed to be the effective flow stress for single slip systems. Compared to the almost ubiquitously adopted Taylor relationship, the derived flow stress formula exhibits two features that are more consistent with the underlying discrete dislocation dynamics: i) the flow stress increases with the in-plane component of the dislocation density only up to a certain value, hence the derived formula admits a minimum inter-dislocation distance within slip planes; ii) the flow stress smoothly transits to zero when all dislocations become geometrically necessary dislocations. A regime under which inhomogeneities in dislocation density grow is identified, and is further validated through comparison with discrete dislocation dynamical simulation results. Based on the findings in this article and in our previous works, a general strategy for incorporating short-range dislocation correlations in continuum models of dislocations is proposed.
\end{abstract}

\begin{keyword}
Continuum model of dislocations \sep Dislocation dipole  \sep  Dislocation pattern formation \sep Flow stress \sep Short-range dislocation correlations


\end{keyword}

\end{frontmatter}

\section{Introduction\label{Sec_introduction}}
It has been widely observed that as crystalline solids undergo plastic deformations, the density distributions of dislocations are generally not homogeneous but form patterns, where dislocation bundles are surrounded by regions occupied by dislocation structures of low density. Such patterned dislocation structures, which form and sometimes transit from one type to another as their matrix specimens are being loaded, play a key role in determining macroscopic materials properties even after unloading. For example, the onset of the persistent slip bands (PSBs) from neighbouring (morphologically different) channel-vein structures marks a different stage in fatigued single-crystals induced by cyclic loads \citep{Mughrabi1976}. For disclosing the links between macroscopic materials properties and their microstructures, a variety of simulation approaches incorporating dislocation dynamics have been developed. For example, ``the state of art'' three-dimensional discrete dislocation dynamics (DDD) simulation tools are developed for capturing dislocation patterns  \citep[e.g.][]{Madec2002, Bulatov_Nature2006,ElAlwardy_NatCom2015}. However, as pointed out by \citet{Sandfeld_pattern2015}, in order to visualise the formation of dislocation patterns in DDD simulations, the computational domains should be at least three times larger than the intrinsic wavelength of the high-density cells of dislocations. Besides, the evolution time generally needs to be long enough to reach certain strain levels under which dislocation patterns are observed experimentally. For example, PSBs normally emerge after thousands of cycles of loads with the strain amplitude of each cycle exceeding a certain point \citep{Mughrabi1976}. To satisfy these requirements, the computational intensity of the resulting DDD simulations is in general too high for most existing DDD simulation tools to afford.

Therefore, it is still highly desirable to develop models of dislocation continua which effectively summarise the underlying discrete dislocation dynamics. In fact, the emergence of continuum models of dislocations dates back to 1950s, when the idea of Nye's dislocation density tensor (abbreviated as the ``Nye's tensor'' in the rest of this article) was proposed \citep{Nye1953, Kroener1958}. In continuum frameworks that only employ the Nye's tensor to represent dislocation structures, the stress field due to the elastic interaction of dislocations is resolved merely in a mean-field sense. One of the crucial aspects that are missing is the short-range elastic correlation between dislocations, which plays a key role in dislocation pattern formations. The reason is in the following. The Nye's tensor is defined by averaging dislocation bundles within some representative volume, and a premise of this averaging process is that the interactions between dislocation segments within the representative volume are negligible at a macroscopic level. This assumption is not quite the case for systems of dislocations, whose motion is generally restricted in their intrinsic slip planes. Because of such anisotropy, various types of \textbf{self-locked dislocation structures} (SLDSs) are formed in crystalline materials. For example, a positive dislocation forms a dipole pair with a negative dislocation residing on a different slip plane, rather than annihilate it. The decoupling stress of dislocation dipoles, which is related to the flow stress for single slip systems, scales (singularly) with the dipole width $r$ by $\mu b/r$ where $\mu$ is the shear modulus and $b$ is the modulus of the Burgers vector. This means that a small change in (discrete) inter-dislocation distance $r$ may bring about considerable variations in the local flow stress. Such position-sensitive feature is also exhibited by many other types of SLDSs appearing in multi-slip systems \citep{Zhu_Scripta2016}. Hence there arises a dilemma. On one hand, models of dislocation continua are expected to be built only in terms of mean-field quantities such as dislocation density fields, while on the other hand, some discrete features of the SLDSs have to be taken into account in order to capture the flow strength. How to reconcile this dilemma becomes the main challenge in developing models of dislocation continua for the past two decades, and no consensus has been reached so far, given a number of valuable continuum frameworks proposed \citep[e.g.][]{Groma1997,ElAzab2000,Acharya2001,Arsenlis2002,Groma2003,Hochrainer2007, Xiang2009_JMPS,Zbib2010,Hochrainer2014,Ngan2015,Katrin_JMPS2015,Geers_JMPS2015, Acharya2015, Zhu_continuum3D,Finel2016,Groma2016,Monavari2016,Ngan_JMPS2016}.

A reasonable starting point for systematic discrete-to-continuum transition is to investigate the collective behaviour of a simple but representative dislocation configuration, where dislocations are straight, mutually parallel and of edge type. In this scenario, the kinematics of dislocation systems is well described by coupled evolution equations for dislocation densities of opposite sign, and the main difficulty lies in how to summarise the ensemble velocities of dislocations (of opposite sign) from the underlying discrete dynamics. One widely adopted mean is to derive the free energy of the dislocation system as a function of dislocation densities, and the dynamics is then formulated either by means of the gradient flow of the free energy \citep[e.g.][]{Groma1999,Groma2006, Zaiser2013,Kooiman2014,Geers_JMPS2015}, or by taking more general trajectories that minimise the free energy \citep[e.g.][]{Mielke2008, Mielke2011,Peletier2015}. Such energetic approaches, however, do not automatically give rise to an expression for the threshold stress value so as to dissociate the SLDSs, i.e. the flow stress, which has to be included additionally either by statistically averaging DDD \citep{Groma2003} or by following conventions \citep[e.g.][]{Schulz2014, Sandfeld_pattern2015}. In most existing continuum models proposed for capturing dislocation pattern formation, two common ingredients are contained. First, the flow stress takes the Taylor relationship \citep{Taylor1934, Franciosi1982,Madec_PRL2002} at its leading order. Second, the average velocities of dislocations of opposite sign are assumed symmetric, i.e. they are of the same magnitude but of opposite sign. These two characteristics enable above mentioned works to take short-range dislocation correlations into account to a certain degree, but there is still essential inconsistency with the underlying DDD which is summarised in the following three aspects.

Firstly, the flow stress given by the Taylor relation $\tau_{\text{f}}\propto\mu b\sqrt{\rho_{\text{tot}}}$  monotonically increases with the total density of dislocations $\rho_{\text{tot}}$. Although such monotonicity gives rise to the growth of inhomogeneities in dislocation densities which is necessary for modelling pattern formation \citep{Sandfeld_pattern2015}, it also results in singular behaviour of the dislocation system. One can imagine that dictated by the Taylor relationship, an SLDS can get infinitely strong by keeping absorbing dislocations. However, it is shown in this article both numerically and analytically that there exists a saturation density point, beyond which the flow stress decreases as the density of the dislocations forming SLDSs increases. Secondly, existing continuum models for dislocation pattern formation are in general inconsistent with the formulations of the dynamics of pure geometrically necessary dislocations (GNDs). It is known that the flow stress in single slip systems when only GNDs are at present should vanish (if the Pierls stress is neglected), and the dislocation velocity is proportional to the local mean-field stress. However, this is not a natural outcome under the set up of Taylor relationship along with the symmetry assumption in dislocation velocities. Note that the limitation of such symmetry assumption is also addressed by the recent works of \citet{Finel2016} and \citet{Groma2016}. The third aspect that limits the existing models from capturing dislocation pattern formation is that they do not resolve the anisotropy in dislocation motion. In most continuum frameworks of dislocations, the short-range dislocation interactions are expressed by functions of local dislocation densities, which are defined as number per area for systems of straight dislocations, and their spatial derivatives. Nevertheless, the rates of change in the two constituent components of the ``number-per-area'' density - the dislocation density within slip planes and the slip plane density, are intrinsically different. In fact, the importance of taking the anisotropy in dislocation motion into account at the continuum level has been addressed in several works. For example, \citet{Geers2013} identified five regimes controlled by the ratio of slip plane spacing to in-plane dislocation spacing in the continuum limit of a regular-wall configuration of GNDs. It is also shown by \citet{Zhu_2Ddipoles2014} that given a same ``number-per-area'' density, dislocation dipoles can form distinct equilibrium patterns.

Therefore, it is more physically significant to derive the expressions of the flow stress and other continuum quantities concurrently with the upscaling of the underlying DDD. One way is to develop hybrid simulation tools \citep{Zbib2010, Acharya2015}, where DDD computational boxes are embedded in a media formed by dislocation continua. As for analytical results that are more applicably favoured, one of the core issues, as stated above, is how to resolve (at the continuum level) the SLDSs induced by the (position-sensitive) short-range dislocation interactions. By studying the collective behaviour of a row of dislocation dipoles, \citet{dipole_SIAP2016} show that the stress field calculated in DDD models can be asymptotically reproduced by the sum of a mean-field stress and a locally periodic discrete stress with the period taking a length scale over which the distribution of dislocation dipoles varies slowly. Based on such stress decomposition, they also show that the collective behaviour of a row of dislocation dipoles can be formulated by a set of evolution equations for dislocation densities along with a quasi-static equilibrium equations for the dipolar width, and the flow stress is then determined by checking the solvability conditions of the equilibrium equations for the dipolar width. Taking a similar strategy, dislocation piling-up effect against various types of dislocation locks due to the interaction of dislocations belonging to non-parallel slip planes is also formulated at the continuum level, based on which a series of flow stress formulae resolving more discrete information than the Taylor relationship are derived \citep{Zhu_Scripta2016}.

In this article, we examine the collective behaviour of arbitrary combinations of straight edge dislocations on two parallel slip planes. The continuum dynamics including the formation, migration and dissociation of dislocation dipoles is formulated in good accordance with DDD simulation results. The contribution of this article is summarised in the following three aspects.

First we propose two continuum velocity field quantities that are more effective for summarising DDD: the local migration velocity of dislocation ensembles and the decoupling velocity of dislocation pairs. It will be shown both numerically and analytically that the local migration velocity of dislocation ensembles, which is the average velocity of all dislocations within a representative volume, is independent of short-range dislocation correlations. Hence its continuum expression can be explicitly derived. It is also shown that, for dislocation ensembles containing GNDs, the dislocation migration velocity is zero only when the external stress is zero. This is actually different from the conventional way of imposing a bifurcating flow condition that the dislocation velocities (of both signs) are zero when the local (mean-field) stress falls below the flow stress. In fact, the bifurcation only emerges in the other proposed velocity field quantity - the decoupling velocity of dislocation pairs, which is defined to be half of the difference between the average velocities of dislocations with opposite signs. The critical stress value beyond which dislocations of opposite sign start to decouple is suggested to be the effective flow stress of single slip systems.

The second contribution of this article is the derivation of an approximate flow stress formula. Compared with the Taylor relationship, the derived formula exhibits two features that are more consistent with DDD: i) the flow stress increases with the in-plane density only up to a certain value, and thus the derived formula admits a minimum inter dislocation distance within slip planes; ii) the flow stress automatically transits to zero as all dislocations become GNDs.

The third contribution of this article is the identification of a regime under which inhomogeneities in dislocation density grow. By numerically investigating the linear instability of the continuum model, we find that a small perturbation to a homogeneous density distribution of dislocations only grows when the following two conditions are satisfied. First, the magnitude of the mean-field stress is greater (and only slightly greater) than the flow stress corresponding to the unperturbed density. Second, this flow stress must fall in the regime where flow stresses increase with the in-plane density of dislocations. The regime of instability identified by the continuum model is validated by DDD simulations. The DDD simulations also suggest that the (initially linear) instability will eventually cease to evolve at a certain stage, and this DDD long-time behaviour is also effectively captured by the derived continuum model.

The continuum model developed in this article is expected to act as a crucial step towards systematically building plasticity theory of dislocation continua, where the dislocation configurations are more complicated. For example, the two introduced velocity quantities can be generalised to facilitate the upscaling of the dynamics of straight dislocations distributed on multiple parallel slip planes. Moreover, one can use the discrete-to-continuum transition approach devised in this article, along with the findings by \citet{Zhu_Scripta2016}, to upscale the dislocation dynamics involving SLDSs induced by the interactions of dislocations belonging to non-parallel slip planes. Based on the findings in the presented and other related works \citep{dipole_SIAP2016, Zhu_Scripta2016}, we summarise a general strategy for incorporating SLDSs into continuum models of dislocations, as outlined in \S~\ref{Sec_discussion}

This article is arranged as follows. Continuum quantities for upscaling discrete dislocation dynamics are introduced in \S~\ref{Sec_d2c}, and their expressions are determined in \S~\ref{Sec_derivation_v}. After listing all the equations constituting the continuum model in \S~\ref{Sec_eqn_summary}, we identify the regime under which a homogeneous distribution of dislocation dipoles is unstable to small perturbations in \S~\ref{Sec_stability}. In \S~\ref{Sec_discussion}, a strategy for taking general SLDSs into account at the continuum level is given, and the article concludes with a summary in \S~\ref{Sec_summary}.

\section{Discrete-to-continuum transition\label{Sec_d2c}}
\subsection{Governing equations at the discrete level}
\begin{figure}[!ht]
  \centering
  \includegraphics[width=.8\textwidth]{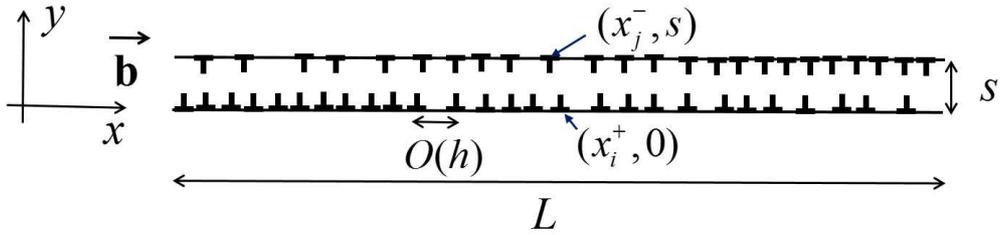}
  \caption{Problem set up.\label{fig_problem_set_up}}
\end{figure}

In this article, we consider straight edge dislocations on two parallel slip planes with the slip plane spacing denoted by $s$ as shown in Fig.~\ref{fig_problem_set_up}.
We here consider the case where there are $N^+$ positive dislocations lying on the slip plane $y=0$ and $N^-$ negative dislocations lying on the slip plane $y=s$. Note that the dislocation sources and annihilation are not taken into account in this work. At $(x_i^+,0)$ locates $i$-th positive dislocation and at $(x_j^-,s)$ locates $j$-th negative dislocation with $1\le i \le N^+$ and $1\le j\le N^-$. Here the computational domain is chosen to be $[-L/2,L/2]$ with $L$ the domain size. The discrete dislocation dynamics is thus described by an ordinary differential equation system:
\begin{subequations}
\beq\label{eqn_DDD_positive0}
\frac1{m_{\text{g}}}\frac{\d x_i^+}{\d t} = \frac{V_i^+}{m_{\text{g}}} = \sum_{j=1}^{N^+}\tau_{\text{ind}}(x^+_i-x_j^+,0) - \sum_{j=1}^{N^-}\tau_{\text{ind}}(x^+_i-x_j^-,s) + \tau_{\text{e}}
\eeq
and
\beq \label{eqn_DDD_negative_start}
\frac1{m_{\text{g}}}\frac{\d x_i^-}{\d t} = -\frac{V_i^-}{m_{\text{g}}} = \sum_{j=1}^{N^+}\tau_{\text{ind}}(x^-_i-x_j^+,s) + \sum_{j=1}^{N^-}\tau_{\text{ind}}(x^-_i-x_j^-,0) - \tau_{\text{e}},
\eeq
\end{subequations}
where $V_i^{+}$ denotes the gliding velocity of $i$-th positive dislocation and so on for $V_i^{-}$; $m_{\text{g}}$ is the dislocation glide coefficient; $\mu$ and $\nu$ are the shear modulus and Poisson's ratio, respectively; $\tau_{\text{e}}$ is the externally applied shear stress; $\tau_{\text{ind}}(\cdot,\cdot)$ formulates the elastic interaction between two individual dislocations of positive sign:
\beq \label{tau_short_single}
\tau_{\text{ind}}(x,y) = \frac{\mu b}{2\pi(1-\nu)}\cdot \frac{x(x^2-y^2)}{(x^2+y^2)^2}.
\eeq

\subsection{Kinematics at the continuum level\label{Sec_kinematics}}
In general, the domain size $L$ is much greater than the average inter-dislocation distance characterised by a parameter $h$, and this results in large dislocation numbers $N^{\pm}\approx L/h$ in the system. Hence it is more efficient to formulate the dislocation dynamics at a continuum level, where the material microstructures are described by the density distributions of dislocations. For configurations consisting of straight and mutually parallel dislocations, dislocation densities are defined as number per area. In this article, the dislocation system is represented by the two components of the ``number-per-area'' density: the slip plane density and dislocation density within slip planes, so as to take dislocation anisotropy into account. Here we fix the slip plane density to be $1/s$, and look for the evolution equation for the in-plane density component, whose definition is associated with a representative interval of length $\delta x$. For example, the in-plane density of positive dislocations at $x$ denoted by $\rho^+(x)$ is defined such that $\rho^+(x)\delta x$ counts the number of positive dislocations within the interval $[x-\delta x/2, x+\delta x/2]$, and so on for the in-plane density of negative dislocations at $x$ denoted by $\rho^-(x)$.
Unless further declared, the term ``dislocation density'' in the rest of this article means ``the density of dislocations within slip planes'', and the input $x$ is dropped for simplicity.

The parameter $\delta x$ controls the resolution of the resulting continuum model. In general,
\beq \label{length_hierarchy}
h \ll \delta x \ll L
\eeq
is required such that the resulting continuum model well trade-off between resolution and computational efficiency.

Note that we only consider the case where the slip plane spacing $s\ll L$. This is because when $s\sim L$, interactions between dislocations on different slip planes can be well resolved by the mean-field stress.

Given $\rho^{\pm}$, the GND density is defined by
\beq \label{density_GND_def}
\rho_{\text{g}} = \rho^{+} - \rho^{-}
\eeq
which may be negative, and the total dislocation density is defined by
\beq \label{density_total_def}
\rho_{\text{tot}} = \rho^{+} + \rho^{-}.
\eeq
The dipole density, which is the number of dislocations forming pairs per unit length, is defined by
\beq \label{density_SSD_def}
\rho_{\text{d}} = \min(\rho^{+},\rho^{-}) = \rho_{\text{tot}} - |\rho_{\text{g}}|.
\eeq

At the continuum level, the dislocation kinematics is described by
\begin{subequations}
\beq \label{eqn_rho_positive}
\pd{\rho^{+}}{t} + \pd{\rho^{+}v^{+}}{x} = 0
\eeq
\beq \label{eqn_rho_negative}
\pd{\rho^{-}}{t} + \pd{\rho^{-}v^{-}}{x} = 0.
\eeq
\end{subequations}
where $v^+$ denotes the average velocity of all positive dislocations in the representative interval centered at $x$ and so on for $v^-$. With Eqs.~\eqref{density_GND_def} and \eqref{density_total_def}, the kinematical relation can be alternatively formulated by the evolution equations for $\rho_{\text{g}}$ and $\rho_{\text{tot}}$:
\beq \label{eqn_density_GND0}
\pd{\rho_{\text{g}}}{t} + \frac1{2}\pd{\rho_{\text{g}}(v^{+}+v^{-})}{x} + \frac1{2}\pd{\rho_{\text{tot}}(v^{+}-v^{-})}{x}= 0;
\eeq
\beq \label{eqn_density_total0}
\pd{\rho_{\text{tot}}}{t} + \frac1{2}\pd{\rho_{\text{g}}(v^{+}-v^{-})}{x} + \frac1{2}\pd{\rho_{\text{tot}}(v^{+}+v^{-})}{x}= 0.
\eeq

\subsection{Continuum formulation of the velocity fields\label{Sec_ddd_set1}}
The continuum velocity fields $v^{\pm}$ are the spatial and time average of the velocities of the discrete dislocations (of either sign) within the representative interval. Mathematically, $v^{\pm}$ at point $x$ and time $t$ (introduced in a continuum sense) are given by
\beq \label{velocity_upscaling}
v^{\pm}(x,t) = \frac1{\delta t}\int_{t-\frac{\delta t}{2}}^{t+\frac{\delta t}{2}} \frac1{n_0^{\pm}}\sum_{i=1}^{n_0^{\pm}} V_i^{\pm} \d t_{\text{dd}},
\eeq
where $n_0^{\pm}$ are the numbers of positive or negative dislocations within the representative interval determined by
\beq \label{determine_n0}
\frac{n_0^+}{n_0^-} \approx \frac{\rho^+}{\rho^-},
\eeq
and the differential $\d t_{\text{dd}}$ indicates that the integration is taken on the time scale associated with the discrete dynamics. The time average in Eq.~\eqref{velocity_upscaling} is made over period $\delta t$, which is chosen much longer than the discrete time scale, such that the quantities after being averaged are expected to depend on mean-field quantities only. However, in the presence of the (position sensitive) short-range elastic interactions of dislocations, finding DDD-consistent expressions for $v^{\pm}$ is extremely difficult. For simplification, two assumptions are widely adopted in existing continuum models. First $v^+=-v^-$. Second, there exits some flow stress value $\tau_{\text{f}}$ below which $v^+=v^-=0$. Here we will show by DDD numerical examples that these two assumptions do not resemble the practical situation in general.

\begin{figure}[!ht]
  \centering
  \subfigure[]{\includegraphics[width=.4\textwidth]{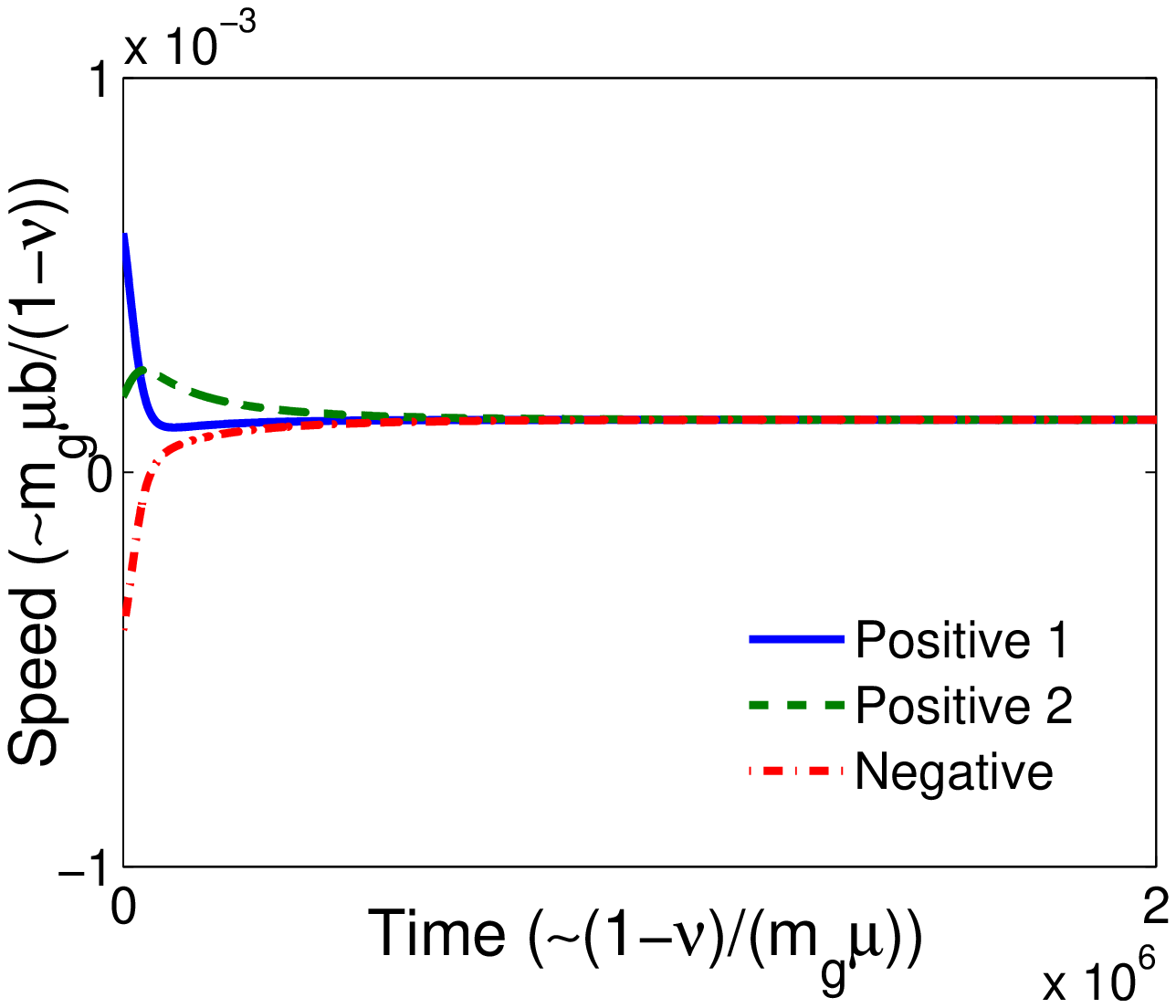}}
  \subfigure[]{\includegraphics[width=.4\textwidth]{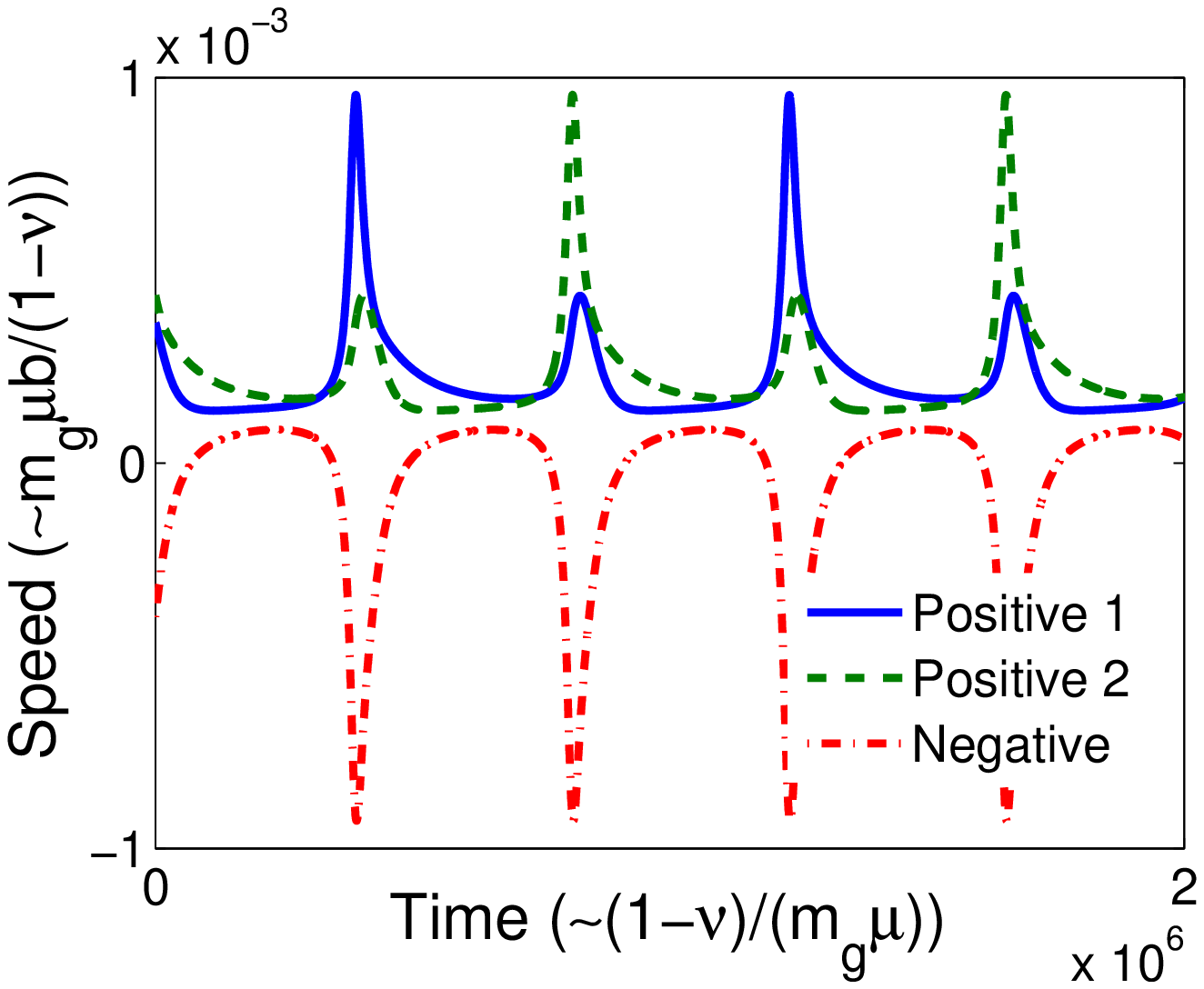}}
  \caption{Behaviour of two positive dislocations and one negative dislocation put in an a representative interval of length $\delta x=500b$ with periodic boundary conditions. The external stress is $\tau_{\text{e}} = 0.004\mu/(1-\nu)$. (a) When $s=75b$, all dislocations are finally bound to drift together at a constant velocity. (b) When $s=100b$, the dislocation pairs get decoupled intermittently. \label{fig_3particles}}
\end{figure}

In a simulation box of length $500b$ which is assumed to be a representative interval, the behaviour of two positive dislocations ($n_0^+=2$) and one negative dislocation ($n_0^-=1$) is studied with periodic boundary conditions, and the result is summarised in Fig.~\ref{fig_3particles}.
When the slip plane spacing $s=75b$, under an external stress of $\tau_{\text{e}}=0.004\mu/(1-\nu)$, all dislocations drift together at a constant velocity after a short period of self-adjustment (Fig.~\ref{fig_3particles}(a)). This is because $\tau_{\text{e}}$ can not overcome the binding stress due to the interaction between dislocations of opposite sign. As all dislocations drift together, $v^{\pm}$ take the same sign, thus $v^+\neq -v^-$. When $s$ is increased to $100b$, the binding force between dislocations becomes weaker, and dislocation pairs get decoupled intermittently (Fig.~\ref{fig_3particles}(b)). We further find that the average velocity of the three dislocations is always proportional to $\tau_{\text{e}}$ (under however small $\tau_{\text{e}}$). This means $v^{\pm}=0$ only takes place when $\tau_{\text{e}}=0$, which is different from the conventional definition of the flow stress.

The DDD simulations also suggest that it is more physically significant to use the following two continuum velocity field quantities other than $v^{\pm}$ to upscale the underlying discrete dynamics: the (local) migration velocity of dislocation ensembles and the average decoupling velocity of dislocation pairs.

The average migration velocity of dislocations denoted by $v_{\text{m}}$ is defined to be the average velocity of all dislocations within the representative interval (averaged over a certain time period $\delta t$):
\beq \label{v_mig_def_ddd}
v_{\text{m}} = \frac1{\delta t}\int_{t-\frac{\delta t}{2}}^{t+\frac{\delta t}{2}} \frac{n_0^+\bar{V}^+ + n_0^-\bar{V}^-}{n_0^+ + n_0^-} \d t_{\text{dd}},
\eeq
where $\bar{V}^+$ is the average velocity of all positive dislocations within the representative interval and so on for $\bar{V}^-$:
\beq \label{velocity_average_pm}
\bar{V}^{\pm} = \frac1{n_0^{\pm}} \sum_{i=1}^{n_0^{\pm}} V_i^{\pm}.
\eeq
Note that $v_{\text{m}}$ is related to $v^{\pm}$ by
\beq \label{v_mig_def_continuum}
\rho_{\text{tot}} v_{\text{m}} = \rho^+v^+ + \rho^-v^-.
\eeq
It will be shown later that $v_{\text{m}}$ is actually independent of the short-range interactions between dislocations.

The second velocity field variable, the decoupling velocity of dislocation pairs denoted by $v_{\text{dc}}$, is defined to be half the average relative velocity between dislocations of opposite sign:
\beq \label{v_sl_def_ddd}
v_{\text{dc}} = \frac{v^+ - v^-}{2} = \frac1{\delta t}\int_{t_0-\frac{\delta t}{2}}^{t_0+\frac{\delta t}{2}} \frac{\bar{V}^+ - \bar{V}^-}{2} \d t.
\eeq
The DDD results presented in Fig.~\ref{fig_3particles} suggest that $v_{\text{dc}}$ bifurcates, and the threshold stress is later identified as the effective flow stress associated with the underlying SLDSs.

Based on Eqs.~\eqref{v_mig_def_continuum} and \eqref{v_sl_def_ddd}, one can express $v^{\pm}$ by
\begin{subequations}
\beq \label{velocity_positive_con}
v^{+}= v_{\text{m}} + \frac{2\rho^- v_{\text{dc}}}{\rho^+ + \rho^-} = v_{\text{m}} + \frac{(\rho_{\text{tot}}-\rho_{\text{g}}) v_{\text{dc}}}{\rho_{\text{tot}}}
\eeq
and
\beq \label{velocity_negative_con}
v^{-}= v_{\text{m}} - \frac{2\rho^+ v_{\text{dc}}}{\rho^+ + \rho^-} =  v_{\text{m}} - \frac{(\rho_{\text{tot}}+\rho_{\text{g}}) v_{\text{dc}}}{\rho_{\text{tot}}},
\eeq
\end{subequations}
where Eqs.\eqref{density_GND_def} and \eqref{density_total_def} are used for obtaining the second identities in both formulae.

Incorporating Eqs.~\eqref{velocity_positive_con} and \eqref{velocity_negative_con} into \eqref{eqn_density_GND0} gives the evolution equation for the GND density:
\beq \label{eqn_density_GND}
\pd{\rho_{\text{g}}}{t} + \pd{\rho_{\text{g}}v_{\text{m}}}{x} + \pd{}{x} \left(\rho_{\text{tot}}v_{\text{dc}} \left(1-\frac{  (\rho_{\text{g}})^2}{(\rho_{\text{tot}})^2} \right)\right)= 0.
\eeq
Similarly, one can incorporate Eqs.~\eqref{velocity_positive_con} and \eqref{velocity_negative_con} into \eqref{eqn_density_total0} to derive the evolution equation for the total density:
\beq \label{eqn_density_total1}
\pd{\rho_{\text{tot}}}{t} + \pd{\rho_{\text{tot}}v_{\text{m}}}{x} = 0,
\eeq
which indicates that the change in $\rho_{\text{tot}}$ is independent of the dissociation process of dislocation dipoles. Eqs.~\eqref{eqn_density_GND} and \eqref{eqn_density_total1} are the governing equations of the continuum model with the two velocity quantities $v_{\text{m}}$ and $v_{\text{dc}}$ to be determined.

It is remarked that one can further employ a dislocation density potential function (DDPF) $\phi(x,t)$, such that $\rho_{\text{g}}=\partial \phi/\partial x$ \citep{Xiang2009_JMPS}. Then Eq.~\eqref{eqn_density_GND} becomes
\beq \label{eqn_DDPF}
\pd{\phi}{t} + v_{\text{m}}\pd{\phi}{x} = - \rho_{\text{tot}}v_{\text{dc}} \cdot \left(1-\frac{  (\rho_{\text{g}})^2}{(\rho_{\text{tot}})^2} \right).
\eeq
A comparison between Eq.~(13) in \citet{Zhu2014_IJP} and Eq.~\eqref{eqn_DDPF} implies that the dissociation of dislocation dipoles serves the same function as active Frank-Read sources.

\section{Expressions of continuum quantities from discrete dislocation dynamics\label{Sec_derivation_v}}
\subsection{Continuum approximation of the stress field\label{Sec_con_approx}}
\begin{figure}
  \centering
  \includegraphics[width=.7\textwidth]{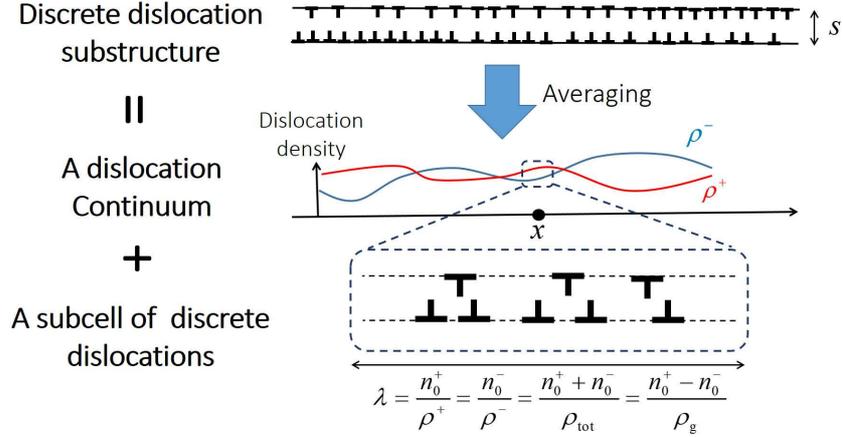}
  \caption{The dislocation system is considered as a dislocation continuum, at each spatial point of which a subcell of discrete dislocations is embedded. The interactions between dislocations inside the subcell are formulated in a discrete sense with periodic boundary conditions imposed. The size of the subcell which is denoted by $\lambda$ satisfies Eq.~\eqref{subdomain_size}. \label{fig_stress_decomposition}}
\end{figure}

Now we search for continuum expressions of quantities defined in \S~\ref{Sec_ddd_set1}: the migration velocity of dislocation ensembles $v_{\text{m}}$ and the decoupling velocity of dislocation pairs $v_{\text{dc}}$. In order to accurately capture (at the continuum level) the stress field due to the position-sensitive short-range dislocation correlations, we introduce a subcell which is associated with each spatial point $x$ (defined at the continuum level), such that the intra-subcell dislocation interactions are still computed in a discrete sense, while the stress due to dislocations outside the subcell is evaluated in a mean-field sense (c.f. Fig.~\ref{fig_stress_decomposition}).
Now the question is how to choose the size of the subcell and the matching conditions at its boundaries such that the calculated stress field is consistent with that evaluated in DDD models. For a row of dislocation dipoles, \citet{dipole_SIAP2016} show that the stress calculated in a DDD sense can be asymptotically re-produced by the sum of a mean-field stress and a locally periodic (discrete) stress with the period taking a length scale on which the SLDSs vary slowly. Based on their findings, one candidate of such locally periodic subcells can be the representative intervals introduced in \S~\ref{Sec_kinematics} with which dislocation densities are defined. With Eq.~\eqref{determine_n0}, the subcell size denoted by $\lambda$ is thus determined by
\beq \label{subdomain_size}
\lambda = \frac{n_0^+}{\rho^+} = \frac{n_0^-}{\rho^-} = \frac{n_0^++n_0^-}{\rho_{\text{tot}}} = \frac{n_0^+-n_0^-}{\rho_{\text{g}}},
\eeq
where $n_0^{\pm}$ are recalled to be the numbers of dislocations of opposite sign inside the representative interval. Here we require $n_0^{\pm}$ to be coprime to each other so as to make the subcell as short as possible.

Then based on the results by \cite{dipole_SIAP2016}, the DDD stress at $i$-th positive dislocation, which is the two discrete sums in Eq.~\eqref{eqn_DDD_positive0}, is approximated by
\beq \label{stress_dis_to_con}
\begin{aligned}
& \quad \sum_{j=1}^{N^+}\tau_{\text{ind}}(x^+_i-x_j^+,0) - \sum_{j=1}^{N^-}\tau_{\text{ind}}(x^+_i-x_j^-,h)\\
& = \sum_{j=1, j\neq i}^{n_0^{+}} \tau_{\text{per}}(x_i^+ - x_j^{+},0) - \sum_{j=1}^{n_0^{-}} \tau_{\text{per}}(x_i^{+} - x_j^{-},s) + \frac{\mu b}{2\pi(1-\nu)} \int_{-L/2}^{L/2} \frac{\rho_{\text{g}}(x')\d x'}{x_i^+-x'},
\end{aligned}
\eeq
where the integral formulates the mean-field stress due to GNDs outside the subcell and $\tau_{\text{per}}(\cdot,\cdot)$ formulates the locally periodic discrete stress with period $\lambda$:
\beq \label{tau_per}
\tau_{\text{per}}(x,y) = \frac{\mu b}{2(1-\nu)\lambda} \cdot \left(\frac{\sin\left(\frac{2\pi x}{\lambda}\right)}{\cosh\left(\frac{2\pi y}{\lambda}\right) - \cos\left(\frac{2\pi x}{\lambda}\right)} - \frac{2\pi y \sin\left(\frac{2\pi x}{\lambda}\right) \sinh\left(\frac{2\pi y}{\lambda}\right)}{\lambda\left(\cosh\left(\frac{2\pi y}{\lambda}\right) - \cos\left(\frac{2\pi x}{\lambda}\right) \right)^2}\right).
\eeq
It can be checked that
\beq \label{tau_per_odd}
\tau_{\text{per}}(-x,y) = -\tau_{\text{per}}(x,y), \qquad \tau_{\text{per}}(x,-y) = \tau_{\text{per}}(x,y).
\eeq
Incorporating Eq.~\eqref{stress_dis_to_con} into \eqref{eqn_DDD_positive0}, we reformulate the (discrete) velocity of $i$-th positive dislocation by
\beq \label{eqn_DDD_positive}
\frac{V_i^{+}}{m_{\text{g}}b} = \sum_{j=1, j\neq i}^{n_0^{+}} \tau_{\text{per}}(x_i^{+} - x_j^{+},0) - \sum_{j=1}^{n_0^{-}} \tau_{\text{per}}(x_i^{+} - x_j^{-},s) + \tau_{\text{m}}(x_i^+),
\eeq
where $\tau_{\text{m}}$ is the mean-field stress which includes the contributions from the long-range interactions of dislocations and the external stress:
\beq \label{tau_mean_field}
\tau_{\text{m}}(x) = \tau_{\text{e}} + \frac{\mu b}{2\pi(1-\nu)} \int_{-L/2}^{L/2} \frac{\rho_{\text{g}}(x')\d x'}{x-x'}.
\eeq
Since the subcell is small viewed at the continuum level, we can assume the mean-field stress $\tau_{\text{m}}$ is a constant within each subcell, and its input $x$ is dropped in later formulations.

Similarly, the discrete velocity of $i$-th negative dislocation is also re-formulated from Eq.~\eqref{eqn_DDD_negative_start}:
\beq\label{eqn_ddd_negative}
\frac{V_i^{-}}{m_{\text{g}}b} = - \sum_{j=1}^{n_0^{+}} \tau_{\text{per}}(x_i^{-} - x_j^{+},s) + \sum_{j=1, j\neq i}^{n_0^{-}} \tau_{\text{per}}(x_i^{-} - x_j^{-},0) -\tau_{\text{m}}.
\eeq

\subsection{Averaging the velocities of discrete dislocations}
In order to summarise for the continuum velocity fields $v_{\text{m}}$ and $v_{\text{dc}}$, we consider the averaged behaviour of all dislocations inside the subcell. If summing Eq.~\eqref{eqn_DDD_positive} for $i=1,2,\cdots, n_0^+$, we obtain
\beq\label{eqn_semi_continuum_sum_positive}
\frac1{m_{\text{g}}b}\sum_{i=1}^{n_0^{+}}V_i^{+} = n_0^{+}\tau_{\text{m}} - \sum_{i=1}^{n_0^{+}}\sum_{j=1}^{n_0^{-}} \tau_{\text{per}}(x_i^{+} - x_j^{-},s).
\eeq
Note that when the summation is made, all terms formulating the interactions between positive dislocations get cancelled because of Eq.~\eqref{tau_per_odd}. Physically, this means the elastic interactions between dislocations are of Newton's third law type.

Dividing both sides of Eq.~\eqref{eqn_semi_continuum_sum_positive} by $n_0^{+}$ gives the instantaneous average velocity of all positive dislocations in the subcell:
\beq\label{velocity_average_positive}
\bar{V}^{+} = m_{\text{g}}b\cdot\left(\tau_{\text{m}} - \frac1{n_0^{+}}\sum_{i=1}^{n_0^{+}}\sum_{j=1}^{n_0^{-}} \tau_{\text{per}}(x_i^{+} - x_j^{-},s)\right).
\eeq
Note that the discrete sum in Eq.~\eqref{velocity_average_positive} formulates the (average) short-range elastic interactions between the dislocations on different slip planes inside the subcell.

Similarly, one can sum Eq.~\eqref{eqn_ddd_negative} over indices $i$ to express the instantaneous average velocity of all negative dislocations in the subcell:
\beq\label{velocity_average_negative}
\bar{V}^{-} = -m_{\text{g}}b\cdot\left(\tau_{\text{m}} - \frac1{n_0^{-}}\sum_{i=1}^{n_0^{+}}\sum_{j=1}^{n_0^{-}} \tau_{\text{per}}(x_i^{+} - x_j^{-},s)\right).
\eeq
Note that the instantaneous average velocity of all dislocations in the subcell denoted by $\bar{V}$ is given by
\beq \label{velocity_average_DDD}
\bar{V} = \frac{n_0^{+}\bar{V}^{+} + n_0^{-}\bar{V}^{-}}{n_0^{+} + n_0^{-}}.
\eeq
Incorporating Eqs.~\eqref{velocity_average_positive} and \eqref{velocity_average_negative} into \eqref{velocity_average_DDD}, we obtain
\beq \label{velocity_migration0}
\bar{V} = \frac{n_0^{+} - n_0^{-}}{n_0^{+} + n_0^{-}}\cdot\tau_{\text{m}} = \frac{\rho_{\text{g}}}{\rho_{\text{tot}}}\cdot m_{\text{g}}b\tau_{\text{m}},
\eeq
where the last identity is based on the relationship between $n_0^{\pm}$ and $\rho^{\pm}$ in Eq.~\eqref{subdomain_size}. Note that the discrete sums in Eqs.~\eqref{velocity_average_positive} and \eqref{velocity_average_negative} cancel with each other again when the summation is made. Eq.~\eqref{velocity_migration0} suggests that $\bar{V}$ is independent of the discrete positions of dislocations, and further the short-range elastic interactions of dislocations. Inserting  Eqs.~\eqref{velocity_average_DDD} and \eqref{velocity_migration0} into \eqref{v_mig_def_ddd}, we obtain the expression of the migration velocity of dislocation ensembles as
\beq \label{velocity_migration1}
v_{\text{m}} = \frac1{\delta t} \int_{t-\delta t/2}^{t+\delta t/2} \frac{m_{\text{g}}b \tau_{\text{m}}\rho_{\text{g}}}{\rho_{\text{tot}}}  \d t_{\text{dd}}.
\eeq
Since $\delta t$ is a short period when viewed at the continuum level, the integrand in Eq.~\eqref{velocity_migration1} which consists of mean-field quantities only, can be treated independent of the discrete time $t_{\text{dd}}$ (but still dependent on continuum time $t$). Hence $v_{\text{m}}$ is expressed by
\beq \label{velocity_migration}
v_{\text{m}} = m_{\text{g}}b \tau_{\text{m}}\cdot \frac{\rho_{\text{g}}}{\rho_{\text{tot}}}.
\eeq
Eq.~\eqref{velocity_migration} implies that $v_{\text{m}}$ is proportional to the product of the mean-field stress and the fraction of GNDs, and this agrees with our observation from DDD simulations shown in Fig.~\ref{fig_3particles}.

Now we look for the continuum expression of $v_{\text{dc}}$. Incorporating Eqs.~\eqref{velocity_average_positive} and \eqref{velocity_average_negative} into \eqref{v_sl_def_ddd} gives
\beq\label{relative_velocity_DDD}
v_{\text{dc}} = \frac{m_{\text{g}}b}{\delta t} \int_{t-\delta t/2}^{t+\delta t/2} \left(\tau_{\text{m}} - \frac{n_0^++n_0^-}{2n_0^+n_0^-} \sum_{i=1}^{n_0^{+}}\sum_{j=1}^{n_0^{-}} \tau_{\text{per}}(x_i^{+} - x_j^{-},s)\right) \d t_{\text{dd}},
\eeq
which suggests that the determination of $v_{\text{dc}}$ requires the resolution of the local discreteness of dislocations.

According to the DDD simulation results presented in Fig.~\ref{fig_3particles}, there should exist a critical stress value, beyond which the decoupling velocity $v_{\text{dc}}$ becomes non-zero. One way to single out this threshold value is to examine the integrand in Eq.~\eqref{relative_velocity_DDD}. If we use $\tau_{\text{f}}$ to denote the maximum value (in an absolute term) the discrete sum in the integrand can reach, i.e.
\beq\label{DDD_fric_max}
\tau_{\text{f}} = \frac{n_0^++n_0^-}{2n_0^+n_0^-} \max_{x_i^{+},x_j^{-}} \left| \sum_{i=1}^{n_0^{+}}\sum_{j=1}^{n_0^{-}} \tau_{\text{per}}(x_i^{+} - x_j^{-},s)\right|,
\eeq
then $\tau_{\text{f}}$ represents the maximum (average) binding stress that can be induced by the short-range elastic interactions between the dislocations on different slip planes. When $|\tau_{\text{m}}|\le\tau_{\text{f}}$, all dislocations in the subcell are able to arrange themselves to generate a short-range stress to counter $\tau_{\text{m}}$. As a result, all dislocations drift at a constant velocity $v_{\text{m}}$ given by Eq.~\eqref{velocity_migration} (as shown in Fig.~\ref{fig_3particles}(a)). When $|\tau_{\text{m}}|>\tau_{\text{f}}$, the stress external to the SLDSs goes beyond the limit of the binding stress, and dislocation pairs start to decouple (Fig.~\ref{fig_3particles}(b)).

The analytical results suggest that a more suitable candidate of the ``effective flow stress'' for single slip hardening is $\tau_{\text{f}}$ given by Eq.~\eqref{DDD_fric_max}, which is the threshold value controlling the bifurcation of the decoupling velocity $v_{\text{dc}}$.

Since $v_{\text{dc}}$ and $\tau_{\text{f}}$ can not be explicitly formulated in terms of the mean-field quantities, we will search for their continuum approximations through upscaling underlying DDD in the next two subsections.

\subsection{Effective flow stress\label{sec_stress_fric_approx}}
\subsubsection{Observation from DDD simulations\label{Sec_ddd_set2}}
DDD simulations are firstly performed so as to reveal the relationship between the flow stress $\tau_{\text{f}}$ and the underlying SLDSs. Similar as the set-up in \S~\ref{Sec_ddd_set1}, a simulation box containing $n_0^+$ positive and $n_0^-$ negative dislocations is loaded under a shear stress component $\tau_{\text{m}}>0$ with periodic boundary conditions. According to its definition, $\tau_{\text{f}}$ is the threshold value of $\tau_{\text{m}}$, when dislocations of opposite sign start to decouple. In the DDD simulations, we start with zero applied shear stress, and $\tau_{\text{f}}$ is determined to be the shear stress when the average velocity difference between dislocations of opposite sign roughly reaches $10^{-3}m_{\text{g}}b\tau_{\text{m}}$. Note that we also need to average in time in DDD simulations.

\begin{figure}[!ht]
  \centering
  \includegraphics[width = .6\textwidth]{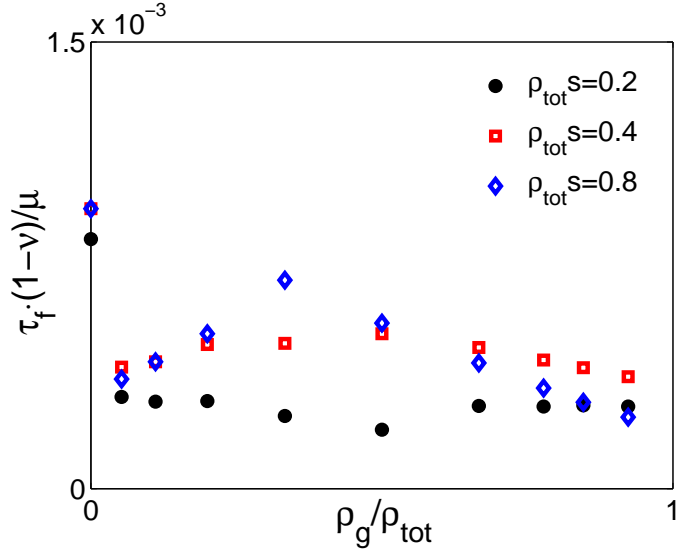}
  \caption{The effective flow stress $\tau_{\text{f}}$ is computed by DDD models with $n_0^+$ positive and $n_0^-$ negative dislocations put in a periodic simulation box. The slip plane spacing is fixed to be $s=50b$. Local dislocation densities are related to $n_0^{\pm}$ based on Eq.~\eqref{subdomain_size}. \label{fig_tau_crit_from_ddd}}
\end{figure}

At the continuum level, $\tau_{\text{f}}$ is expected to be a function of $\rho_{\text{g}}$ and $\rho_{\text{tot}}$, whose relationship with $n_0^{\pm}$ is formulated by Eq.~\eqref{subdomain_size}. Given a fixed slip plane spacing $s$, $\tau_{\text{f}}$ is plotted against $\rho_{\text{g}}/\rho_{\text{tot}}$ for various choices of $\rho_{\text{tot}}$ in Fig.~\ref{fig_tau_crit_from_ddd}.
A key observation is that the value of $\tau_{\text{f}}$ drops significantly as $\rho_{\text{g}}$ turns slightly non-zero. The reason behind this drop can be better perceived by referring to the study by \cite{Zhu_Scripta2016}, where the role of dislocation piling-up in multi-slip hardening is formulated at the continuum level. In the scenario of single slip systems here, a non-vanishing $\rho_{\text{g}}$ implies that not all dislocations are locked in dipoles. Under an external load, those excessive dislocations are driven towards their locked neighbours. This piling-up process potentially generates a strong back stress at the locked dislocations, which allows the break-up of a dipole lock under a relatively low external load. Therefore, for any given slip plane spacing $s$, it is always the configurations consisting of pure dislocation dipoles ($\rho_{\text{g}}=0$) that deliver the strongest flow stress, and this phenomenon has also been mentioned in other works \citep[e.g.][]{Groma2010}.

Fig.~\ref{fig_tau_crit_from_ddd} indicates that $\tau_{\text{f}}$ depends discretely on the dislocation density. Thus expressing $\tau_{\text{f}}$ as a continuous function of mean-field variables becomes extremely difficult. If a compromise has to be made, it is more sensible to let the resulting flow stress formula to accurately capture the strongest $\tau_{\text{f}}$ for any $s$. Hence we take the following two steps to derive a continuum flow stress formula. We first search for an accurate description of $\tau_{\text{f}}$ when $\rho_{\text{g}}=0$. Then we approximate the expression of $\tau_{\text{f}}$ for other cases by smoothly connecting the two extreme cases where flow stress are known: $\rho_{\text{g}}=0$ and $\rho_{\text{d}}=0$.

\subsubsection{Effective flow stress in a GND-free state\label{Sec_flow_stress_dipole}}
When all dislocations form dipoles, the (locally periodic) subcell contains only a pair of dislocations, i.e. $n_0^+=n_0^-=1$. Then the flow stress formula \eqref{DDD_fric_max} is simplified to be
\beq \label{tau_crit_dipole0}
\left.\tau_{\text{f}}\right|_{\rho_{\text{g}}=0} = \frac1{2}\max_{x^+,x^-}|\tau_{\text{per}}(x^{+} - x^{-},s)|.
\eeq
For the ease of further analysis, we introduce a non-dimensional function $G$ of two entries, such that
\beq \label{tau_per_G}
\tau_{\text{per}}(x^{+} - x^{-},s) = \frac{\mu b}{4(1-\nu)\lambda}\cdot G\left(\frac{x^{+} - x^{-}}{\lambda},\frac{s}{\lambda}\right),
\eeq
where $\lambda$ is recalled to be the subcell size and according to Eq.~\eqref{subdomain_size} it is given by
\beq \label{subdomain_size_dipole}
\lambda = \frac2{\rho_{\text{tot}}}.
\eeq
A comparison between Eq.~\eqref{tau_per} and \eqref{tau_per_G} implies
\beq \label{G_def}
G(w_1,w_2) = \frac{\sin\left(2\pi w_1\right)}{\cosh\left(2\pi w_2\right) - \cos\left(2\pi w_1\right)} - \frac{2\pi w_2 \sin\left(2\pi w_1\right) \sinh(2\pi w_2)}{\left(\cosh\left(2\pi w_2\right) - \cos\left(2\pi w_1\right) \right)^2}.
\eeq
Incorporating Eqs.~\eqref{tau_per_G} and \eqref{subdomain_size_dipole} into \eqref{tau_crit_dipole0} gives
\beq \label{tau_crit_dipole}
\left.\tau_{\text{f}}\right|_{\rho_{\text{g}}=0} = \frac{\mu b\rho_{\text{tot}}}{4(1-\nu)}\max_{x^+,x^-}\left|G\left(\frac{\rho_{\text{tot}}(x^{+} -x^{-})}{2},\frac{\rho_{\text{tot}}s}{2}\right)\right| = \frac{\mu b\rho_{\text{tot}}}{4(1-\nu)}\max_{z}\left|G\left(z,\frac{\rho_{\text{tot}}s}{2}\right) \right|,
\eeq
where the last identity is obtained by treating $\rho_{\text{tot}}(x^+-x^-)/2$ as a single free variable for optimisation. Eq.~\eqref{tau_crit_dipole} indicates that $\tau_{\text{f}}$ can be expressed as the product of two quantities:
\beq \label{tau_crit_dipole_f}
\left.\tau_{\text{f}}\right|_{\rho_{\text{g}}=0} = \frac{\mu b}{(1-\nu)s}\cdot f(\rho_{\text{tot}}s),
\eeq
where the non-dimensional function $f$ is given by
\beq \label{f_expression}
f(\rho_{\text{tot}}s) = \frac{\rho_{\text{tot}}s}{4} \cdot \max_{z}\left|G\left(z,\frac{\rho_{\text{tot}}s}{2}\right)\right|.
\eeq
Eq.~\eqref{tau_crit_dipole_f} implies that i) $\tau_{\text{f}}$ scales with the slip plane density $1/s$; ii) the dependence of $\tau_{\text{f}}$ on the in-plane dislocation density is mediated by a non-dimensional quantity $\rho_{\text{tot}}s$, which is actually the ratio of the total in-plane dislocation density to the slip plane density.

The complicated form of $G(\cdot,\cdot)$ (given by Eq.~\eqref{G_def}) makes it still too difficult to analytically determine $f(\rho_{\text{tot}}s)$. Instead we fit for the expression of $f(\rho_{\text{tot}}s)$ against data obtained by numerically solving Eq.~\eqref{f_expression}.

\begin{figure}[!ht]
  \centering
  \includegraphics[width=.55\textwidth]{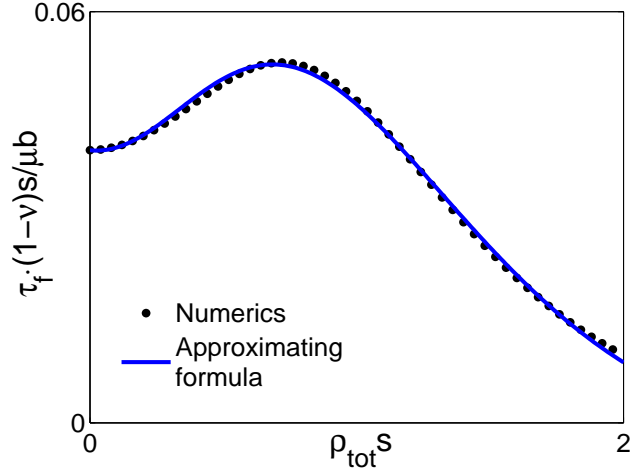}
  \caption{Flow stress $\left.\tau_{\text{f}}\right|_{\rho_{\text{g}}=0}$ corresponding to the configurations consisting of pure dislocation dipoles against $\rho_{\text{tot}}s$ is generated with reference to Eq.~\eqref{tau_crit_dipole_rhos}. The discrete dots are obtained by numerically solving Eq.~\eqref{f_expression}. The curve also corresponds to $f(\rho_{\text{tot}}s)$ given by Eq.~\eqref{f_rhoss}. \label{fig_tau_c_rho_dipole}}
\end{figure}

Asymptotic behaviour of $f(\rho_{\text{tot}}s)$ is investigated first. When $\rho_{\text{tot}}\rightarrow0$, the subcell size $2/\rho_{\text{tot}}$ tends to infinity, which means all dislocation dipoles are well separated. In this case, the flow stress is actually the maximum binding stress a positive dislocation at the origin can exert on the slip plane given by $y=s$, and it can be calculated that $\tau_{\text{f}}=\mu b/(8\pi(1-\nu)s)$. A comparison with Eq.~\eqref{tau_crit_dipole_f} suggests that
\beq \label{f_lower_lim}
\lim_{\rho_{\text{tot}}\rightarrow 0} f(\rho_{\text{tot}}s) = \frac1{8\pi}.
\eeq
On the other hand, when $\rho_{\text{tot}}\rightarrow\infty$, $G(z,\rho_{\text{tot}}s/2)$ in Eq.~\eqref{f_expression} decays like (power of) hyperbolic functions. Hence the target function of $f(\rho_{\text{tot}}s)$ should contain terms carrying this feature. It turns out that $f(\rho_{\text{tot}}s)$ can be approximated by
\beq \label{f_rhoss}
f(\rho_{\text{tot}}s) = \frac{(a_0 + a_1(\rho_{\text{tot}}s)^2)(1+a_2\sinh(a_3\rho_{\text{tot}}s))}{ \cosh^2(a_3\rho_{\text{tot}}s)}.
\eeq
With reference to Eq.~\eqref{f_lower_lim}, $a_0=1/(8\pi)$. The other parameters $a_1=0.1447$, $a_2=-0.0892$ and $a_3=1.2252$ are determined by fitting against numerical data. Incorporating Eq.~\eqref{f_rhoss} into \eqref{tau_crit_dipole_f} gives an approximating formula of the effective flow stress corresponding to the configurations consisting of dislocation dipoles:
\beq \label{tau_crit_dipole_rhos}
\left.\tau_{\text{f}}\right|_{\rho_{\text{g}}=0} = \frac{\mu b}{(1-\nu)s}\cdot\frac{(1/(8\pi) + a_1(\rho_{\text{tot}}s)^2)(1+a_2\sinh(a_3\rho_{\text{tot}}s))}{ \cosh^2(a_3\rho_{\text{tot}}s)}.
\eeq
Eq.~\eqref{tau_crit_dipole_rhos} is compared with the numerical data in Fig.~\ref{fig_tau_c_rho_dipole}, and good agreement is observed for $\rho_{\text{tot}}s<2$.
In this paper, $\left.\tau_{\text{f}}\right|_{\rho_{\text{g}}=0}$ is given by Eq.~\eqref{tau_crit_dipole_rhos} before it turns negative and $\left.\tau_{\text{f}}\right|_{\rho_{\text{g}}=0}$ is zero otherwise. For a more accurate description of $\left.\tau_{\text{f}}\right|_{ \rho_{\text{g}}=0}$ when $\rho_{\text{tot}}s>2$, one way is to employ another target function (for $\rho_{\text{tot}}s>2$).

It is observed from Fig.~\ref{fig_tau_c_rho_dipole} that, given any slip plane spacing, there exists a saturation density value, at which $\tau_{\text{f}}$ is maximised. This is different from the widely used Taylor-type flow stress formulae ($\tau_{\text{f}}\propto\mu b\sqrt{\rho_{\text{tot}}}$), under which an SLDS can get infinitely strengthened by keeping absorbing dislocations from nearby.

The non-monotonicity of $\left.\tau_{\text{f}}\right|_{\rho_{\text{g}}=0}$ with $\rho_{\text{tot}}$ can be attributed to the sign switch in the shear stress field due to a single dislocation. For a row of positive dislocations, their capability of capturing a negative dislocation on the slip plane $y=s$ (which is effectively the flow stress), depends on the strongest shear stress (in absolute terms) they can exert on $y=s$. Suppose there are two positive dislocations as shown in Fig.~\ref{fig_dipole_strength}:
\begin{figure}[!ht]
  \centering
  \subfigure[]{\includegraphics[width=.4\textwidth]{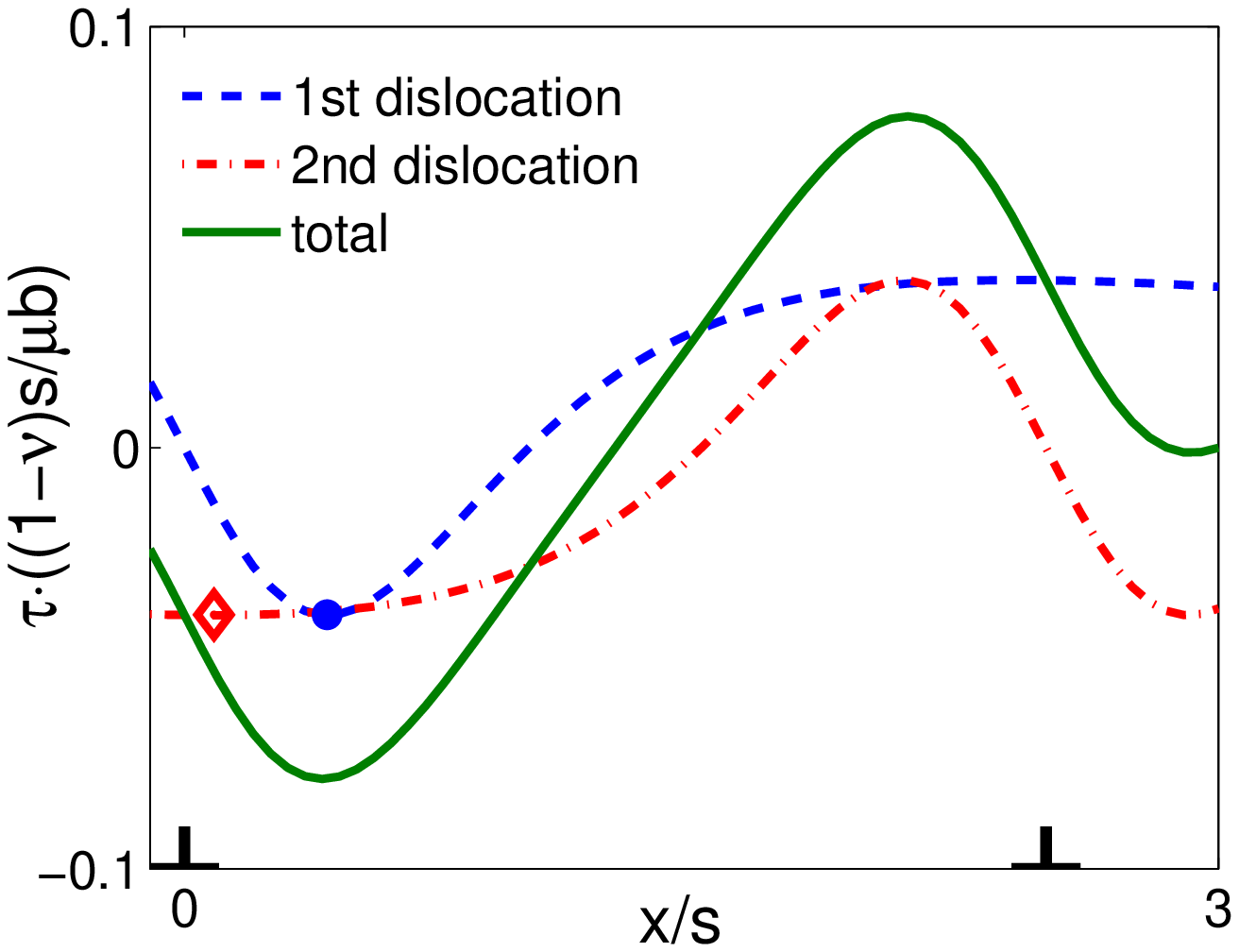}}
  \subfigure[]{\includegraphics[width=.4\textwidth]{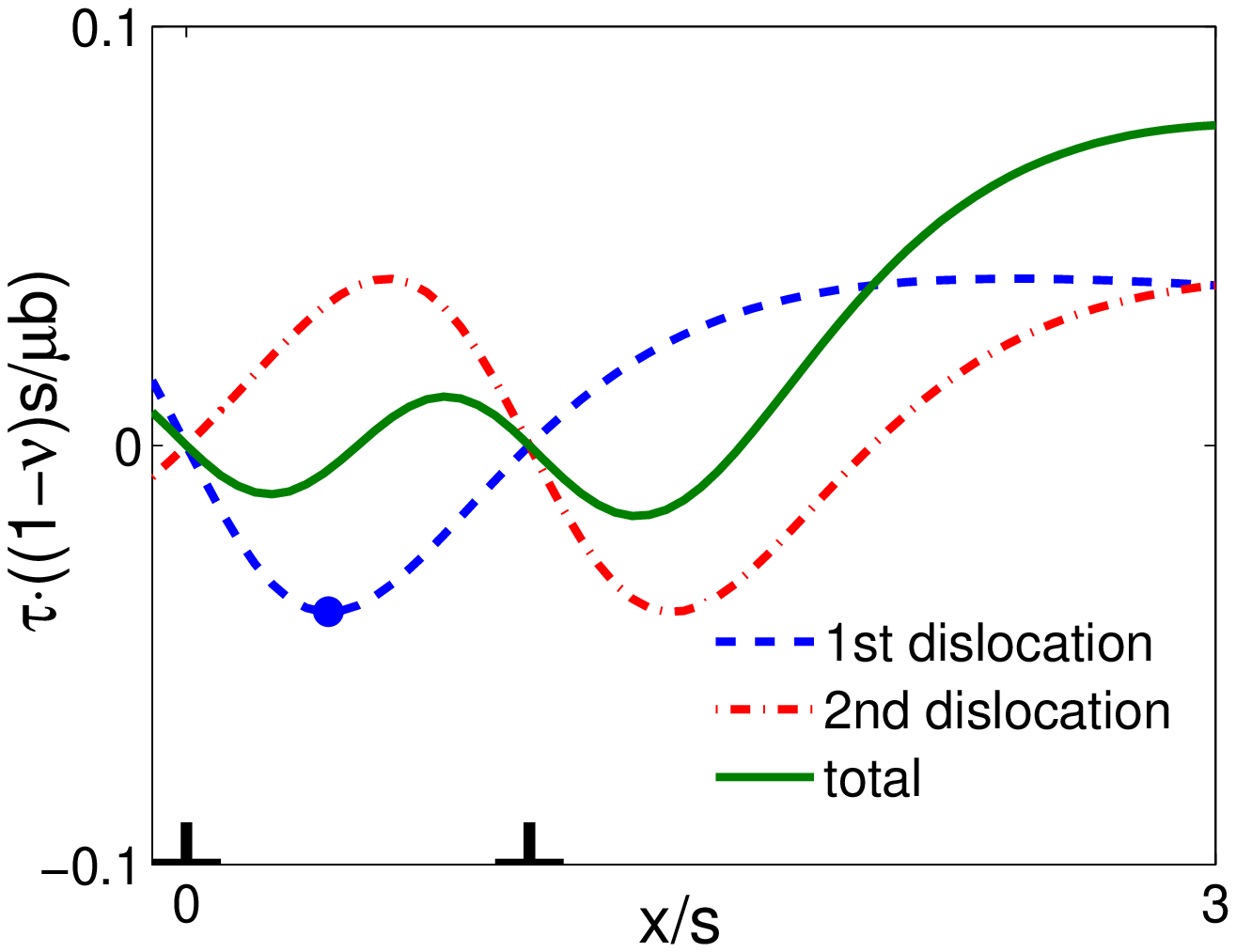}}
  \caption{The total shear stress field on $y=s$ due to two positive dislocations located on $y=0$. The shear stress fields due to individual dislocations are denoted by the blue dashed curves and the red dashed-dotted curves. (a) When the two dislocations are far apart, the total binding stress (the green curve) get strengthened (compared to that of a single dislocation). (b) When the two dislocations are close to each other, the binding stress gets weakened. \label{fig_dipole_strength}}
\end{figure}
one is located at the origin and it exerts a shear stress field on $y=s$ denoted by the blue dashed curves; the other is located at $(x_0,0)$ with its shear stress on $y=s$ denoted by the red dashed-dotted curves. When the two dislocations are far apart (Fig.~\ref{fig_dipole_strength}(a)), the total binding stress (given by the green curve) get strengthened (compared to that of a single dislocation), because some of the negative-stress regions of the two individual dislocations overlap. When the two dislocations are close to each other (Fig.~\ref{fig_dipole_strength}(b)), the negative region of one positive dislocation overlaps with the positive region of the other, resulting in a weakened binding stress. Therefore, the strongest binding stress by two positive dislocations should be roughly attained when the two (negative) peaks (denoted by the blue dot and the red diamond in Fig.~\ref{fig_dipole_strength}(a)) coincide with each other. In this case, the spacing between positive dislocations is calculated to be $2\sqrt{2}s$.

\subsubsection{Non-vanishing GND density}
Then we look for a general expression of $\tau_{\text{f}}$. We assume that the effective flow stress takes the following form:
\beq \label{tau_crit_general0}
\tau_{\text{f}} = p(\rho_{\text{g}},\rho_{\text{tot}})\cdot\left.\tau_{\text{f}}\right|_{ \rho_{\text{g}=0}},
\eeq
where $\tau_{\text{f}}|_{ \rho_{\text{g}=0}}$ is given by Eq.~\eqref{tau_crit_dipole_rhos}. A natural condition for the function $p$ is that $p=1$ when $\rho_{\text{g}}=0$. On the other hand, if all dislocations are GNDs, i.e. $|\rho_{\text{g}}|=\rho_{\text{tot}}$, the effective flow stress vanishes, and this delivers the condition on the other end: $p=0$ as $|\rho_{\text{g}}|/\rho_{\text{tot}}=1$. By letting $p=(1-|\rho_{\text{g}}|/\rho_{\text{tot}})^2$ which satisfies the above conditions, we obtain an approximating formula of the effective flow stress:
\beq \label{tau_crit_general}
\tau_{\text{f}} = \frac{\mu b}{s(1-\nu)}\cdot\left(1-\left|\frac{\rho_{\text{g}}}{\rho_{\text{tot}}} \right|\right)^2\cdot \frac{(a_0 + a_1(\rho_{\text{s}}s)^2)(1+a_2\sinh(a_3\rho_{\text{s}}s))}{ \cosh^2(a_3\rho_{\text{s}}s)},
\eeq
where $a_0=1/(8\pi)$, $a_1=0.1447$, $a_2=-0.0892$ and $a_3=1.2252$.

Note that there are infinitely many choices for the function $p$. In fact, as long as $p$ is assumed continuous with its entries, it is almost impossible to use a formula taking the form of Eq.~\eqref{tau_crit_general0} to capture all the information about local discreteness revealed by the DDD simulation results in \S~\ref{Sec_ddd_set2}. However, formula \eqref{tau_crit_general} carries two basic features of $\tau_{\text{f}}$ that is consistent with underlying discrete dynamics. First, it accurately predicts the flow stress corresponding to configurations consisting of dislocation dipoles, which deliver the strongest flow stress for any given slip plane spacing. Second, it automatically vanishes when all dislocations are GNDs, thus delivers a smooth transition from a non-GND state to a non-dipole state in $\tau_{\text{f}}$.

\subsection{Decoupling velocity of dislocations of opposite sign\label{Sec_determine_v_dc}}
Now we consider formulating the decoupling velocity between dislocations of opposite sign $v_{\text{dc}}$ which should meet two conditions. First, $v_{\text{dc}}=0$ when the magnitude of the mean-field stress $|\tau_{\text{m}}|\le\tau_{\text{f}}$. Second, when $|\tau_{\text{m}}|\gg\tau_{\text{f}}$, the binding force between dislocations of opposite sign becomes negligible, which delivers $v_{\text{dc}}\approx m_{\text{g}}b\tau_{\text{m}}$. Hence we can assume $v_{\text{dc}}$ takes the  following form:
\beq \label{v_sl_tau_general}
v_{\text{dc}} = m_{\text{g}}b\tau_{\text{m}}\cdot \left(1-\left|\frac{\tau_{\text{f}}}{\tau_{\text{m}}}\right|^{\alpha}\right)
\eeq
for $|\tau_{\text{m}}|>\tau_{\text{f}}$ and is zero otherwise. In principle, the parameter $\alpha$ should be determined through comparison with DDD simulation results. Here for simplicity we choose $\alpha=1$.

\subsection{Higher-order stress correction}
One assumption underlying the derived continuum model is that the dislocation density should be slowly varying on some small length scales compared to the specimen size, such that the discrete stress can be well approximated by a mean-field stress and a locally periodic discrete stress. However, when the dislocation density varies sharply in space, this assumption may not well resemble the practical situation as the higher-order stress correction in the above-mentioned stress decomposition becomes non-negligible. For the case of a row of dislocation dipoles, \citet{dipole_SIAP2016} show that this correction stress, which tends to homogenise the distribution of dislocation dipoles, takes the order of magnitude $\mu b/L$ where $L$ is the specimen size. It is a high-order term when compared to the flow stress which $\sim\mu b/h$ with $h$ recalled to be the average inter-dislocation distance. A natural way to include this stress correction here is to introduce a high-order derivative of $\rho_{\text{d}}$ to the equation for the total density $\rho_{\text{tot}}$:
\beq \label{eqn_density_total}
\pd{\rho_{\text{tot}}}{t} + \pd{\rho_{\text{tot}}v_{\text{m}}}{x} = D_0\cdot\frac{\partial^2 \rho_{\text{d}}}{\partial x^2}
\eeq
with the coefficient $D_0$ given by
\beq \label{diffusion_coefficient}
D_0 = \frac{c_0m_{\text{g}}\mu b^2}{1-\nu},
\eeq
where $c_0$ is a non-dimensional parameter. For the case of a row of dislocation dipoles, $c_0$ is found non-linearly dependent on a number of factors, such as the non-dimensional quantity $\rho_{\text{tot}}s$ (for details see Eq.~(5.16) in \citet{dipole_SIAP2016}), and its value roughly takes the order of magnitude 0.1. In the more general scenario as in this article, $c_0$ is chosen to be $0.4$, under which the continuum models show good agreement with the underlying DDD. The reason that we use the high-order derivative of the dipole density $\rho_{\text{d}}$ instead of $\rho_{\text{tot}}$ is for consistency with existing continuum models of GNDs, which will be discussed in detail in \S~\ref{Sec_consistency}. Note that although $D_0$ only depends on materials constants, another internal length scale is introduced by the high-order term in Eq.~\eqref{eqn_density_total}.

\section{Governing equations at the continuum level\label{Sec_eqn_summary}}
\subsection{Equation list}
To summarise, the continuum model of the dynamics of straight dislocations on two parallel slip planes constitutes
\begin{itemize}
  \item an evolution equation for the GND density:
  \beq \label{summary_rhog}
  \pd{\rho_{\text{g}}}{t} + \pd{\rho_{\text{g}}v_{\text{m}}}{x} + \pd{v_{\text{dc}}}{x} \left(\rho_{\text{tot}}v_{\text{dc}} \left(1-\frac{  (\rho_{\text{g}})^2}{(\rho_{\text{tot}})^2} \right)\right)= 0;
  \eeq
  \item an evolution equation for the total density:
  \beq \label{summary_rhot}
  \pd{\rho_{\text{tot}}}{t} + \pd{\rho_{\text{tot}}v_{\text{m}}}{x} = \frac{c_0m_{\text{g}}\mu b^2}{1-\nu}\cdot\frac{\partial^2 \rho_{\text{d}}}{\partial x^2};
  \eeq
  \item a law of dislocation migration velocity:
  \beq \label{summary_vm}
  v_{\text{m}} = m_{\text{g}}b\tau_{\text{m}}\cdot \frac{\rho_{\text{g}}}{\rho_{\text{tot}}};
  \eeq
  \item a law of decoupling velocity of dislocation pairs:
  \beq \label{summary_vrl}
  v_{\text{dc}} = \left\{\begin{aligned}
  & m_{\text{g}}b(\tau_{\text{m}} - \tau_{\text{f}}\cdot\text{sign}(\tau_{\text{m}})), \qquad && |\tau_{\text{m}}| > \tau_{\text{f}};\\
  & 0, && |\tau_{\text{m}}| \le \tau_{\text{f}}.
  \end{aligned}
  \right.
  \eeq
  where the effective flow stress $\tau_{\text{f}}$ is given by Eq.~\eqref{tau_crit_general}.
\end{itemize}

Here we also list the equations for $\rho^{\pm}$ which are equivalent to Eqs.~\eqref{summary_rhog} and \eqref{summary_rhot}:
\begin{subequations}
\beq \label{eqn_rho_positive_numerics}
\pd{\rho^+}{t} + \pd{\rho^+v^+}{x} = \frac{c_0m_{\text{g}}\mu b^2}{2(1-\nu)}\cdot\spd{\rho_{\text{d}}}{x}
\eeq
and
\beq \label{eqn_rho_negative_numerics}
\pd{\rho^-}{t} + \pd{\rho^-v^-}{x} = \frac{c_0m_{\text{g}}\mu b^2}{2(1-\nu)}\cdot\spd{\rho_{\text{d}}}{x},
\eeq
\end{subequations}
where $v^+$ and $v^-$ are determined by Eqns.~\eqref{velocity_positive_con} and \eqref{velocity_negative_con}, respectively. The numerical results to be presented in the next section are obtained based on Eqs.~\eqref{eqn_rho_positive_numerics} and \eqref{eqn_rho_negative_numerics}.

\subsection{Consistency with the well-formulated cases\label{Sec_consistency}}
Here we consider two cases that have been properly formulated in literature, and we will show that they are simply two special cases of the derived continuum model. The first case is when all dislocations are GNDs, e.g. $\rho^+ = \rho_{t} = \rho_{\text{g}}$. The dynamics is known formulated by \citep{Head1972c}
\beq \label{eqn_GND_only}
\pd{\rho^+}{t} + m_{\text{g}}b\cdot\pd{\rho^+\tau_{\text{m}}}{x} = 0.
\eeq
Under the same circumstance, Eq.~\eqref{summary_vm} becomes $v_{\text{m}}=m_{\text{g}}b\tau_{\text{m}}$, and the equation for $\rho_{\text{g}}$ by Eq.~\eqref{summary_rhog} is found taking the same form as Eq.~\eqref{eqn_GND_only}. On the other hand, the high-order derivative in Eq.~\eqref{summary_rhot} vanishes as $\rho_{\text{d}}=0$, and the equation for $\rho_{\text{tot}}$ also takes the form as Eq.~\eqref{eqn_GND_only}.

The other special case considered is when $|\tau_{\text{m}}|\gg\tau_{\text{f}}$. The short-range dislocation interactions can be neglected, and all dislocations stay in a ``plasma state''. The governing equations become \citep{Groma1997}
\beq \label{eqn_plasma}
\pd{\rho^{\pm}}{t} + \pd{\rho^{\pm}v^{\pm}}{x} = 0,
\eeq
where $v^+=-v^-=m_{\text{g}}b\tau_{\text{m}}$. In this scenario, Eq.~\eqref{summary_vrl} becomes $v_{\text{dc}}\approx m_{\text{g}}b \tau_{\text{m}}$, meaning the decoupling velocity of dislocation dipoles does not depend on the effective flow stress. Then referring to Eqs.~\eqref{velocity_positive_con} and \eqref{velocity_negative_con}, we obtain $v^+=-v^-=m_{\text{g}}b\tau_{\text{m}}$. Note that as $|\tau_{\text{m}}|\gg\tau_{\text{f}}$, the higher-order stress correction formulated by the right hand side of Eq.~\eqref{summary_rhot} is also negligible.

\section{Instability in homogeneous distributions of dislocation dipoles \label{Sec_stability}}
\subsection{Short-time instability}
The short-time instability of the derived continuum model is studied first. The regime where small perturbations to homogeneous density distributions of dislocations grow is identified with reference to simulation results based on the continuum model. Here we are interested in the instability conditions when all dislocations form dipoles. For obtaining the simulation results shown later, Eqs.~\eqref{eqn_rho_positive_numerics} and \eqref{eqn_rho_negative_numerics} are numerically evolved. The long-range elastic interactions between dislocations which are formulated by the integral term in Eq.~\eqref{tau_mean_field}, are computed by using the fast Fourier transform (FFT). The Fourier coefficient of $\exp(2\im \pi kx/L)$ is given by $-\im \pi \text{sign}(k)\mathcal{F}\left[\rho_{\text{g}}\right](k)$, where $\mathcal{F}\left[\rho_{\text{g}}\right](k)$ denotes the Fourier transform of $\rho_{\text{g}}$ in $[-L/2,L/2]$: $\mathcal{F}[\rho_{\text{g}}](k) = \frac1{L}\int_{-L/2}^{L/2} e^{-2\im\pi kx/L}\rho_{\text{g}}(x) \d x$.

The first set of numerical examples are directed to show that instability only emerges when the flow stress increases with local dislocation density. In a simulation box of length $L$, we put in $N$ pairs of dislocations. This corresponds to a uniform density $\rho^{\pm}=N/L$, and we perturb it to obtain the initial conditions for our simulations:
\beq \label{rho_initial}
\rho^{\pm} = \left(1- \epsilon \sin \left(\frac{2\pi x}{L}\right)\right)\cdot\frac{N}{L},
\eeq
where the initial amplitude $\eps$ is chosen to be $\pi/20$ for the simulation results presented later.

\begin{figure}[!ht]
  \centering
  \subfigure[Continuum]{
  \includegraphics[width=.4\textwidth]{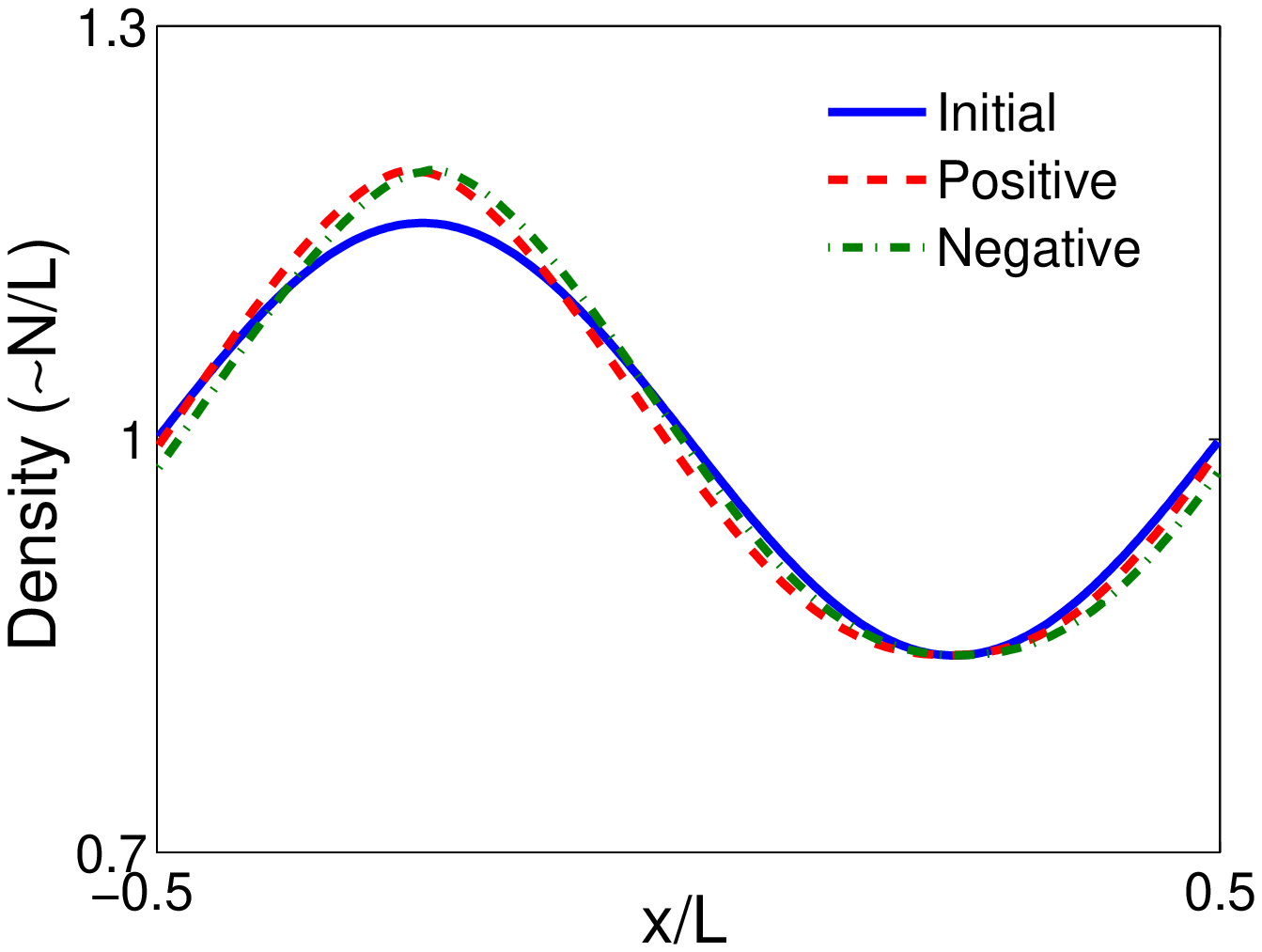}}
  \subfigure[DDD]{
  \includegraphics[width=.4\textwidth]{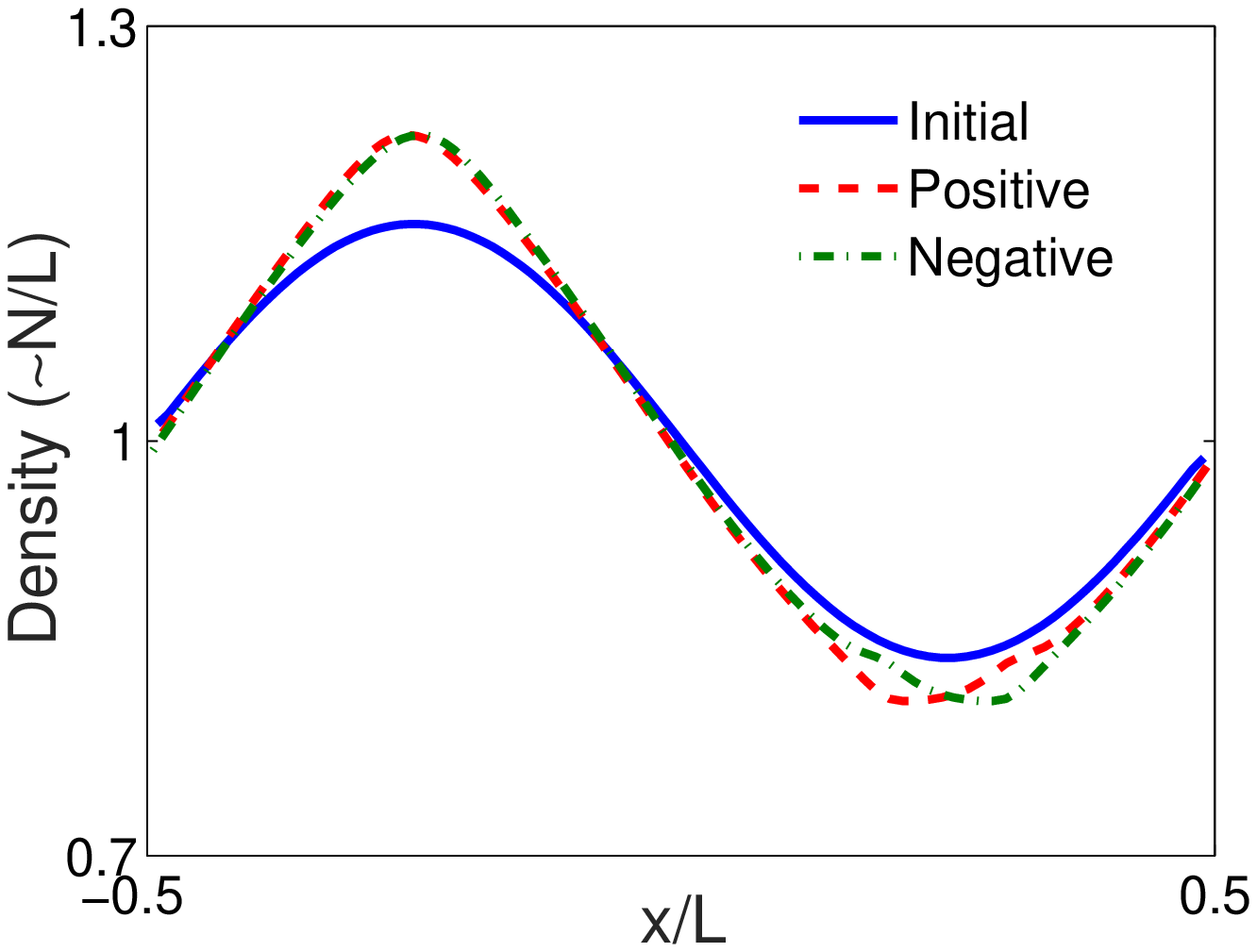}}
  \caption{Instability in homogeneous density to small perturbations: both simulations start from Eq.~\eqref{rho_initial} (given by the solid curves for dislocations of both sign) which is obtained by slightly perturbing a uniform distribution of dislocation dipoles $N/L$. Here the computational domain size $L=5\times10^4b$, the slip plane spacing $s=100b$, the total number of dislocation pairs $N=100$ and the external shear stress $\tau_{\text{e}}=0.24\mu bN/((1-\nu)L)$. The small perturbations in dislocation densities of opposite sign grow in time as suggested by both the continuum and DDD simulations. The density distributions of dislocation of opposite sign (given by the dash-dotted and dashed curves) are taken at $t=5(1-\nu)L^2/(m_{\text{g}} b^2N)$. \label{fig_instability}}
\end{figure}

In the first numerical example, $s=100b$, $N=100$ and $L=5\times10^4b$. With periodic boundary conditions, we generate Fig.~\ref{fig_instability}(a) which suggests the growth of instability.
DDD simulations under same conditions deliver similar outcomes as shown in Fig.~\ref{fig_instability}(b). Note that dislocation densities in DDD simulations are approximated by $\rho^{\pm}\left(\frac{x_i^{\pm}+x_{i-1}^{\pm}}{2} \right)\approx1/(x^{\pm}_i-x^{\pm}_{i-1})$.

\begin{figure}[!ht]
  \centering
  \subfigure[]{
  \includegraphics[width=.4\textwidth]{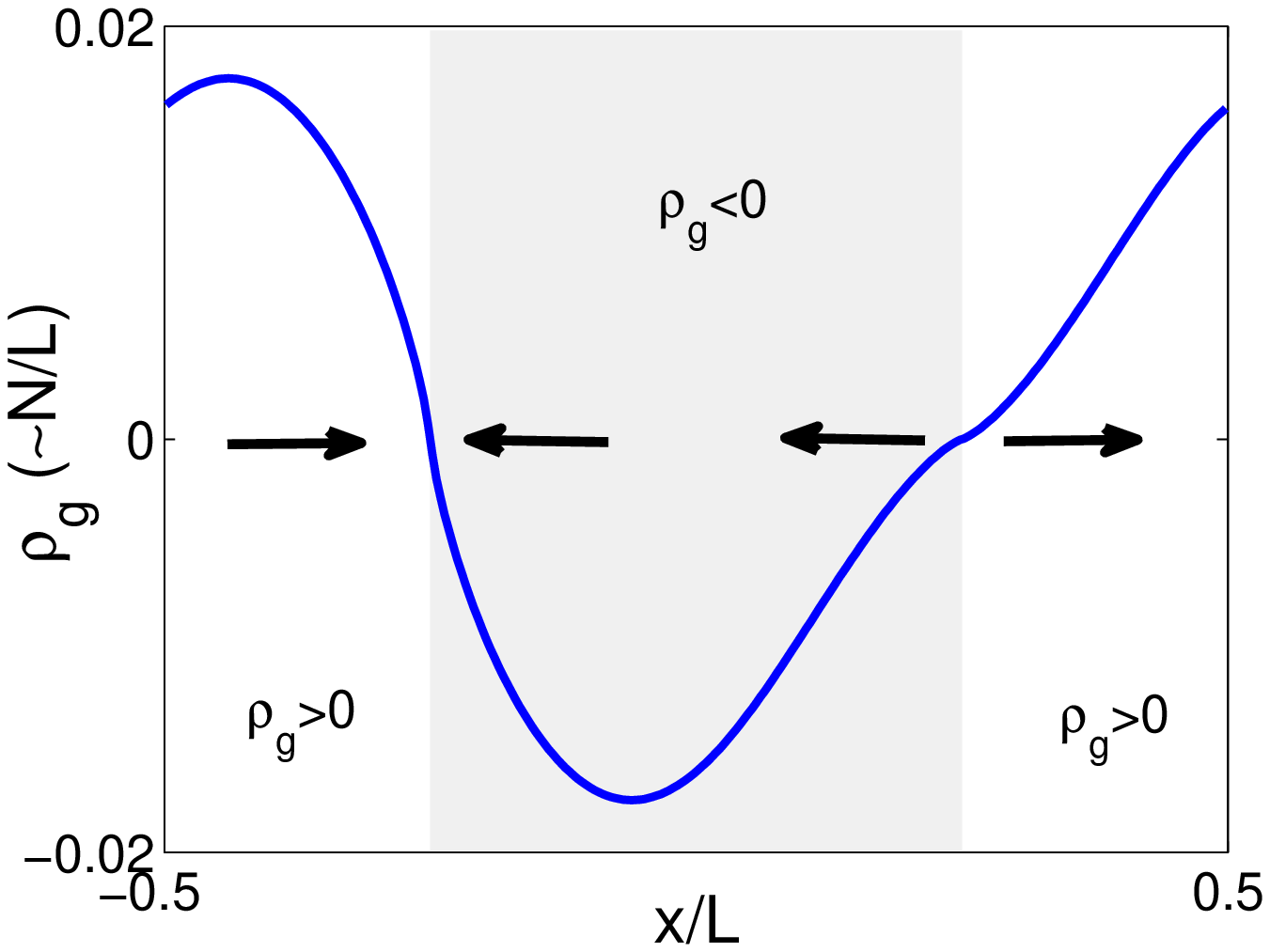}}
  \subfigure[]{
  \includegraphics[width=.4\textwidth]{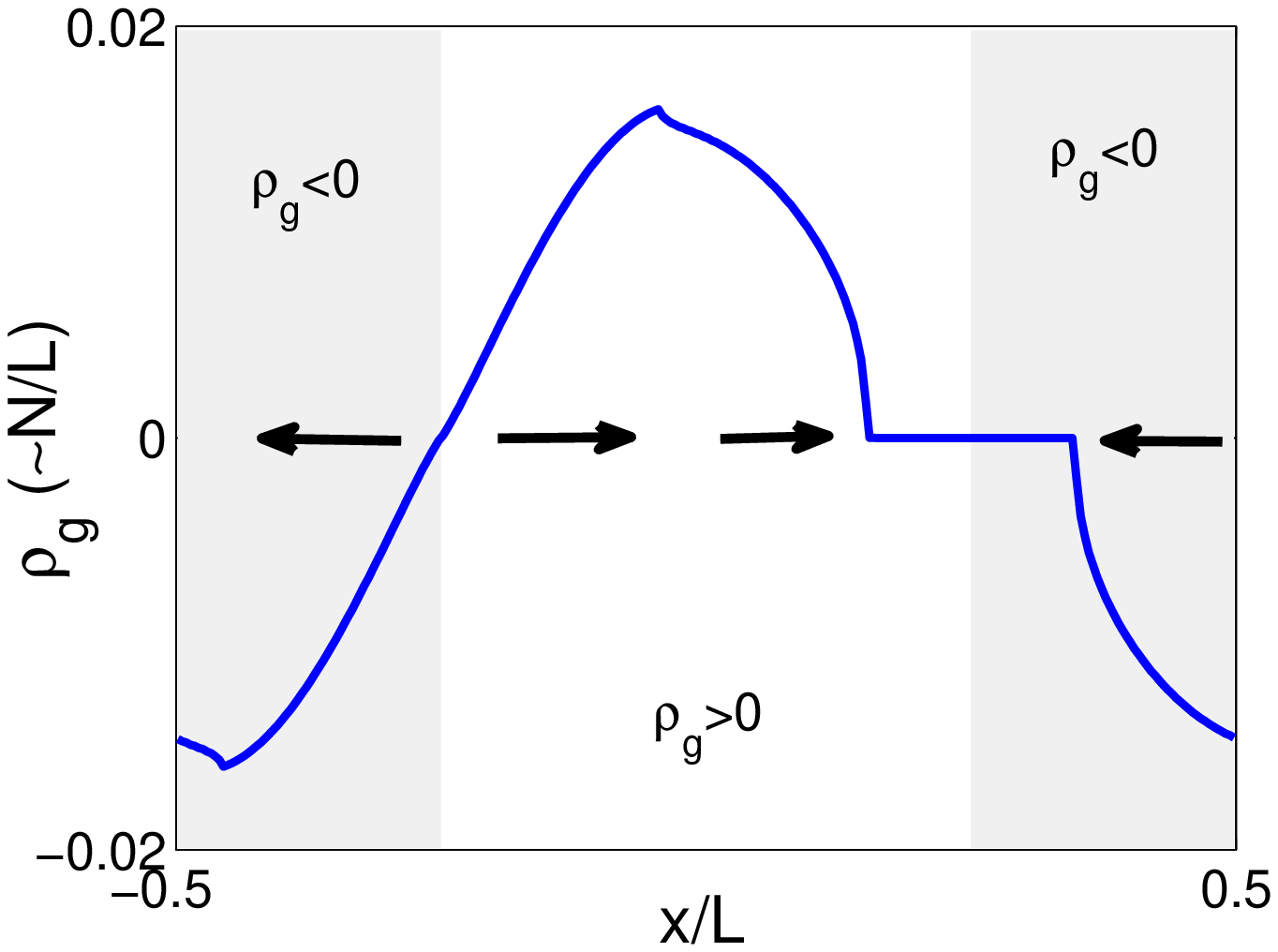}}
  \caption{When the external shear stress $\tau_{\text{e}}$ is slightly higher than the flow stress associated with the dipole structures before being perturbed, dislocation dipoles with weaker flow stress are more likely to be dissociated to GNDs. In response to $\tau_{\text{e}}$, the generated GNDs of opposite sign flow along different directions to regions with higher flow stress and form dipoles there. (a) When $\rho_{\text{tot}}s<1/\sqrt{2}$, the weaker flow stress is taken in the lower-density region centered at $x=0.25L$, and the density inhomogeneities grow as a result; (b) when $\rho_{\text{tot}}s> 1/\sqrt{2}$, the weaker flow stress is taken in the higher-density region centered at $x=-0.25L$, and the system evolves to a homogeneous state. \label{fig_rhog}}
\end{figure}

To rationalise this linear instability, we refer to the expression of $\tau_{\text{f}}$ in Eq.~\eqref{tau_crit_dipole_rhos}, which reads $\tau_{\text{f}}\propto\mu b/s\cdot f(\rho_{\text{tot}}s)$ with $f(\cdot)$ given by Eq.~\eqref{f_rhoss}. In the first numerical example, the non-dimensional quantity $\rho_{\text{tot}}s\approx0.4$ falls in the interval where $f(\cdot)$ is an increasing function. (Recall that $f(x)$ peaks as $x=1/\sqrt{2}$ c.f. Fig.~\ref{fig_tau_c_rho_dipole}.) This means SLDSs of higher density are stronger. Under an external stress slightly above the flow stress corresponding to $\rho_{\text{tot}}s=0.4$, more dislocation dipoles in the low-density region (centered at $x=0.25L$) get dissociated into GNDs than those in the high-density region (centered at $x=-0.25L$). Under $\tau_{\text{e}}>0$, the generated positive GNDs (in the white region in Fig.~\ref{fig_rhog}(a)) travel to the right, while the negative GNDs (in the shaded region in Fig.~\ref{fig_rhog}(a)) flow to the left.
GNDs of opposite sign re-combine to form dipoles in the high-density region. As a result, dislocation density in the originally high-density region increases further, which induces an even larger flow stress value. This is the reason of the linear instability observed in the first numerical example.

\begin{figure}[!ht]
  \centering
  \subfigure[Continuum]{
  \includegraphics[width=.4\textwidth]{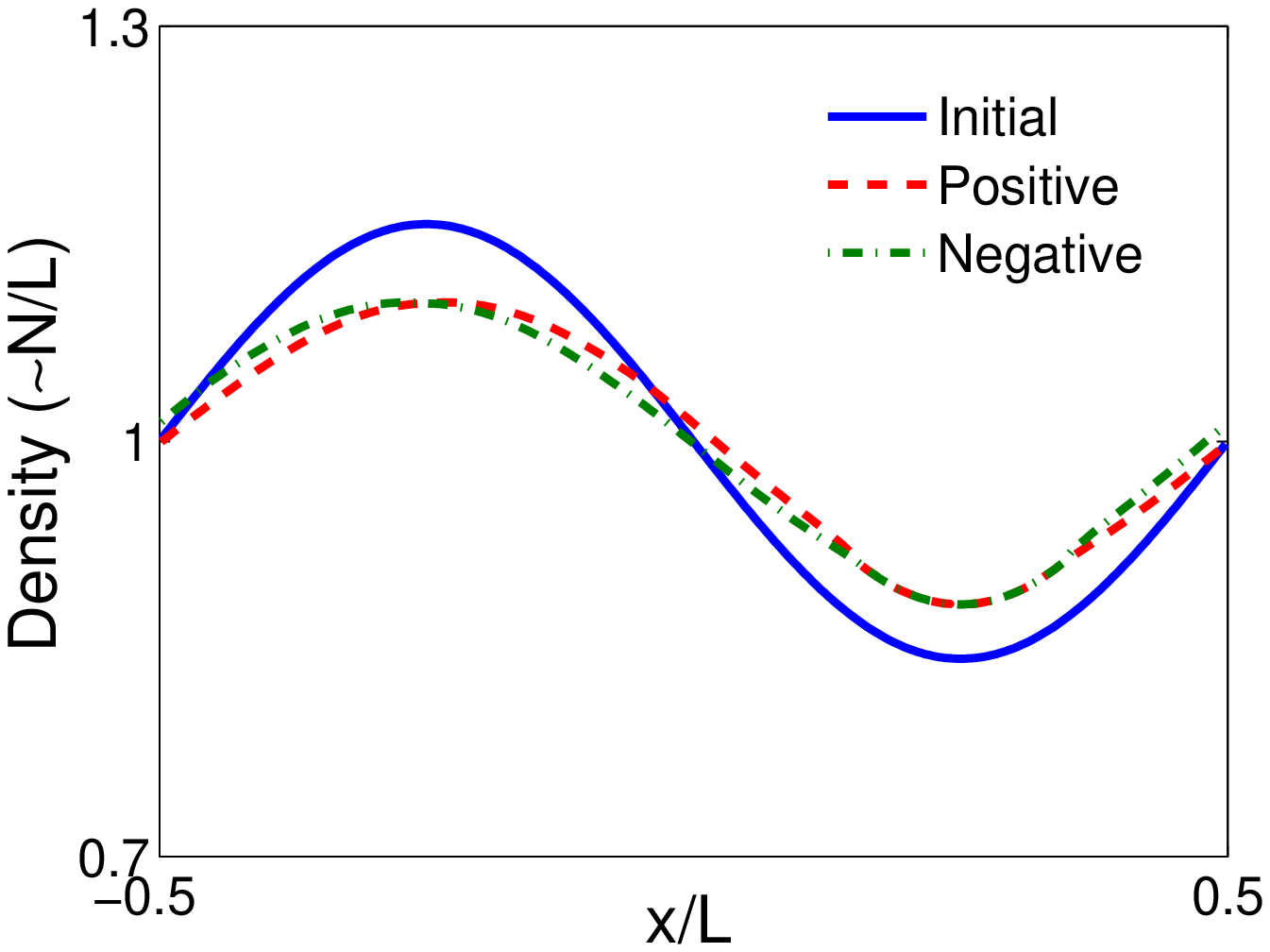}}
  \subfigure[DDD]{
  \includegraphics[width=.4\textwidth]{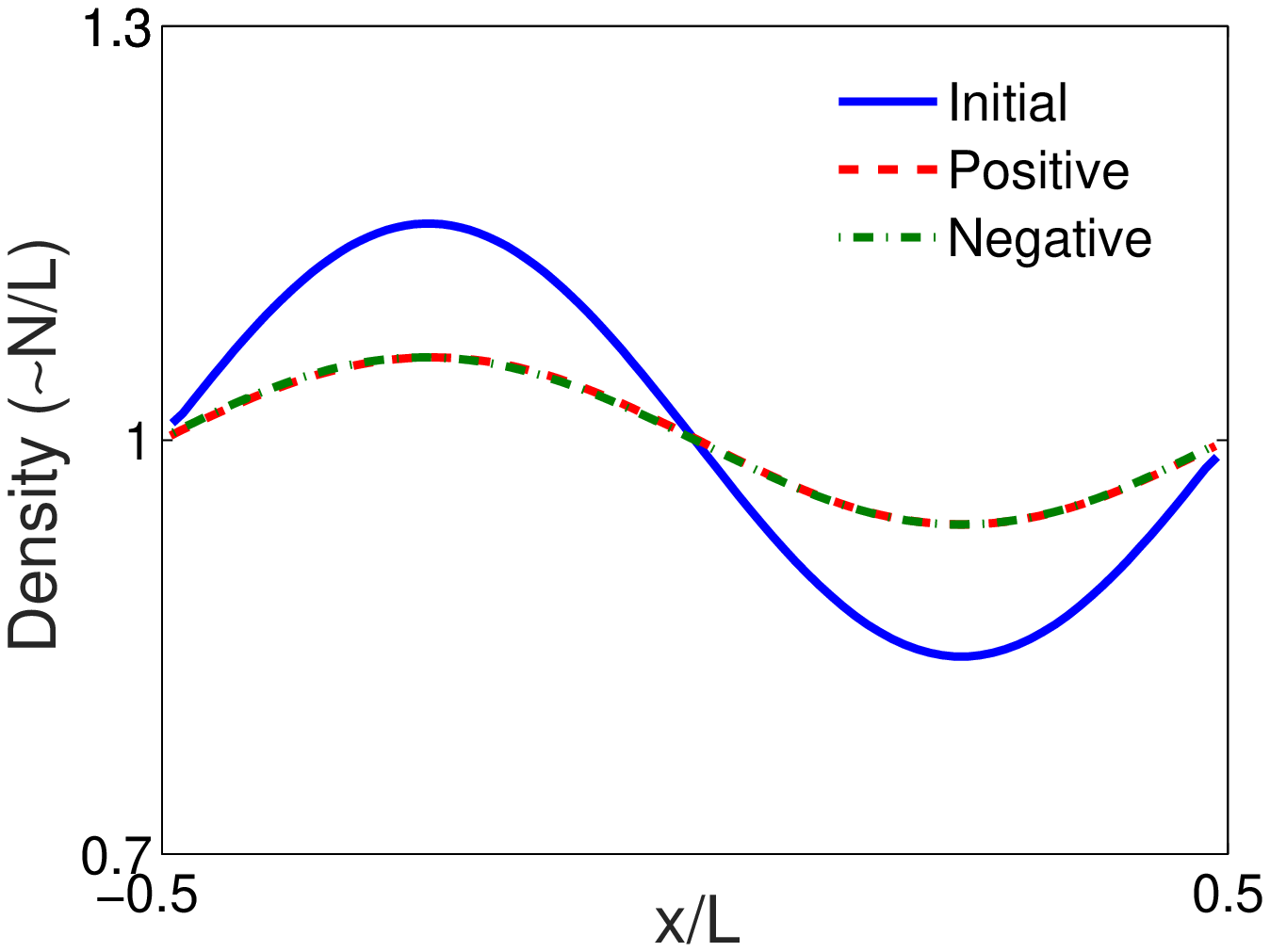}}
  \caption{Starting with Eq.~\eqref{rho_initial}, the small perturbation in dislocation densities decays when $\rho_{\text{tot}}s>1/\sqrt{2}$. Here $L=2\times10^4b$, $s=100b$, $N=100$ and $\tau_{\text{e}}=0.09\mu bN/((1-\nu)L)$. Both continuum and DDD simulations stop at $t=5(1-\nu)L^2/(m_{\text{g}} b^2N)$. \label{fig_stable}}
\end{figure}

In contrast, if we fix $s=200b$ and $N=100$ but shrink $L$ to $2\times10^4b$, $\rho_{\text{tot}}s=1$ which falls into the interval where $f(\cdot)$ is a decreasing function (c.f. Fig.~\ref{fig_tau_c_rho_dipole}). From Eq.~\eqref{tau_crit_dipole_rhos}, the flow stress (corresponding to $\rho_{\text{tot}}s=1$) is computed to be $\tau_{\text{f}}=0.09\mu bN/((1-\nu)L)$. Under $\tau_{\text{e}}=\tau_{\text{f}}$, the small perturbation given by Eq.~\eqref{rho_initial} is found decaying as shown in Fig.~\ref{fig_stable}(a), which agrees with the DDD simulation results shown in Fig.~\ref{fig_stable}(b).
In this case, dislocation dipoles in the high-density region (centered at $x=-0.25L$) get dissociated into GNDs, which recombine to form dipoles in the low-density region (centered at $x=0.25L$), as shown in Fig.~\ref{fig_rhog}(b).

The numerical results above suggest that, under an external stress slightly above the flow stress corresponding to the unperturbed dislocation configurations, inhomogeneities in dislocation densities only grow within a certain range which is given by $\rho_{\text{tot}}s<1/\sqrt{2}$.

\begin{figure}[!ht]
  \centering
  \subfigure[Continuum]{
  \includegraphics[width=.4\textwidth]{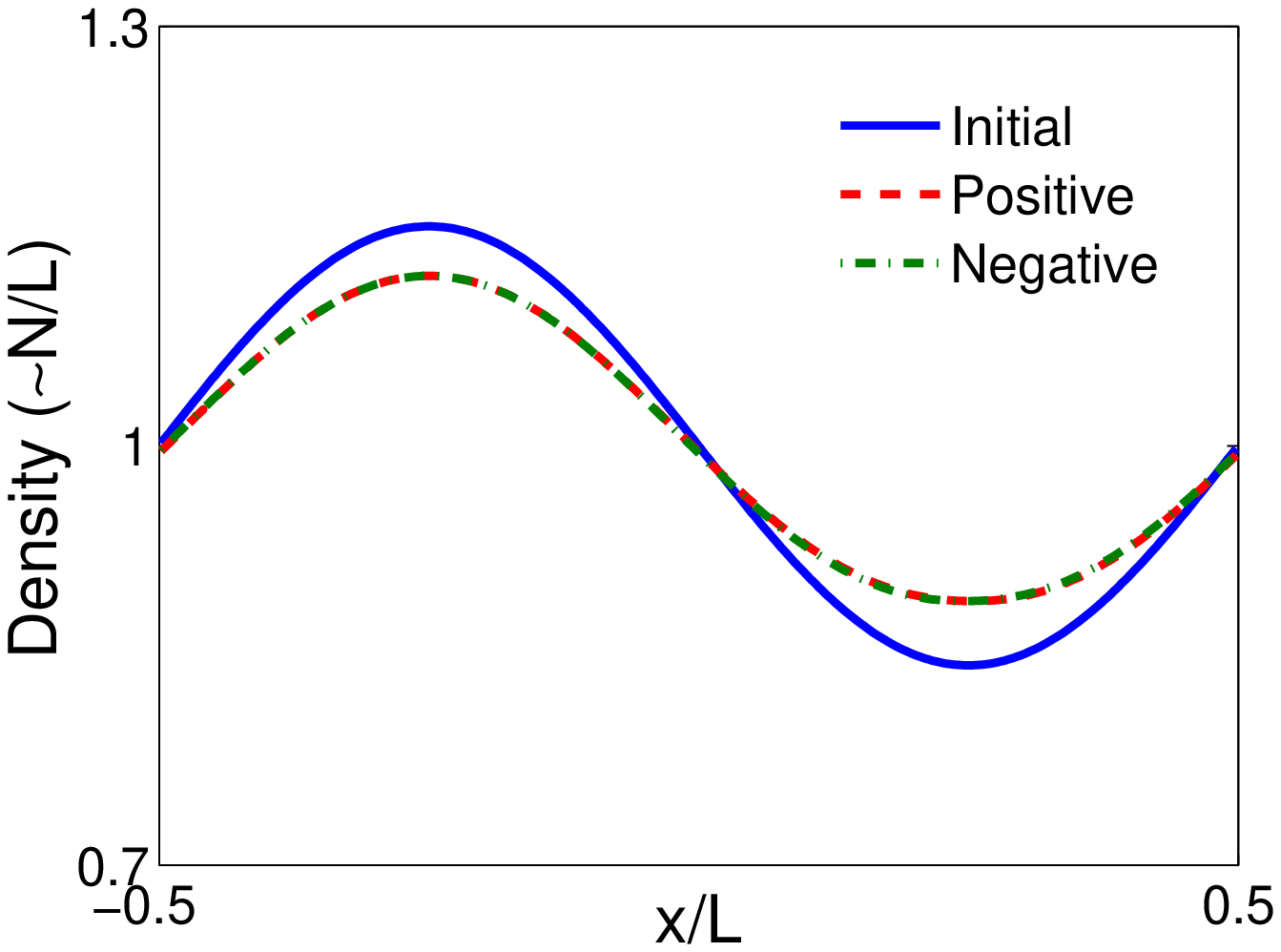}}
  \subfigure[DDD]{
  \includegraphics[width=.4\textwidth]{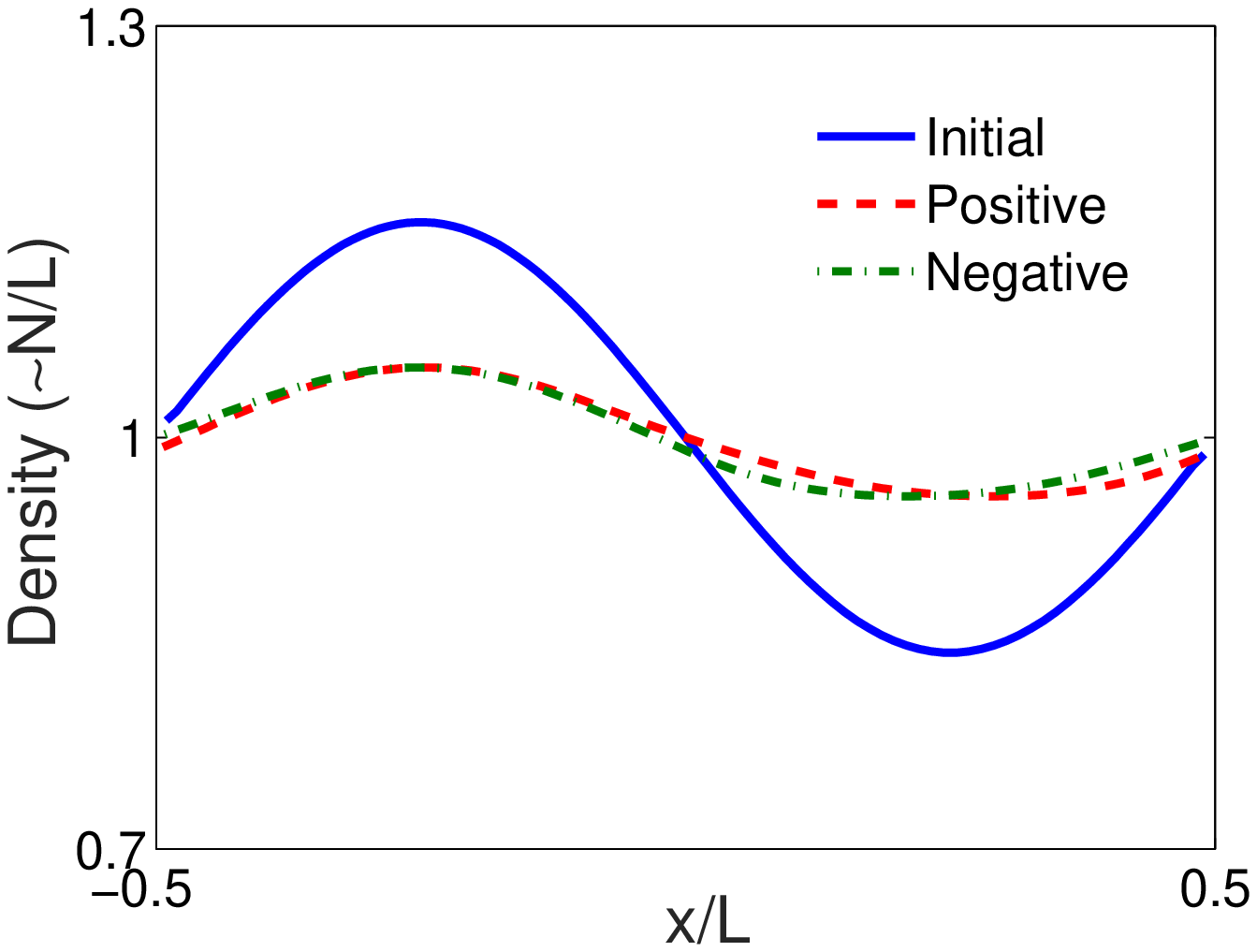}}
  \caption{The small perturbation decays. Here the coefficients are chosen the same as the unstable case given by Fig.~\ref{fig_instability} ($L=5\times10^4b$, $s=100b$, $N=100$) except that $\tau_{\text{e}}=0.3\mu bN/((1-\nu)L)$, which is considerably larger than the flow stress associated with the unperturbed density ($\tau_{\text{f}}\approx0.24\mu bN/((1-\nu)L)$). Both continuum and DDD simulations stop at $t=5(1-\nu)L^2/(m_{\text{g}} b^2N)$. \label{fig_large_tau}}
\end{figure}

For the second set of numerical examples, we keep all simulation conditions the same as obtaining Fig.~\ref{fig_instability} ($L=5\times10^4b$, $s=100b$, $N=100$), except that $\tau_{\text{e}}=0.3\mu bN/((1-\nu)L)$ which is considerably larger than the flow stress of the unperturbed structures ($\tau_{\text{f}}\approx0.24\mu bN/((1-\nu)L)$). Fig.~\ref{fig_large_tau}(a) shows that the initial perturbation decays in this case, which agrees with the DDD simulation results presented in Fig.~\ref{fig_large_tau}(b).
A rationalisation of this regime of stability is that when the difference $|\tau_{\text{e}}|-\tau_{\text{f}}$ becomes more apparent, the dislocation dynamics behaves more like the case where dislocations of opposite sign transport independently. The corresponding evolution equations are given by Eq.~\eqref{eqn_plasma}, which are stable to small perturbations. The finding indicates that dislocation dipoles can hardly sustain at shear stress that is considerably higher than their induced flow stress.

\begin{figure}[!ht]
  \centering
  \includegraphics[width=.55\textwidth]{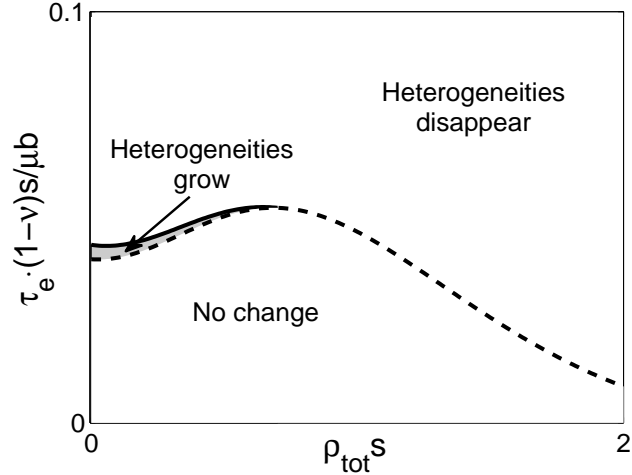}
  \caption{A schematic summary of the regimes of instability suggested by the continuum model. When $|\tau_{\text{e}}|$ falls below $\tau_{\text{f}}$ which is given by the dashed curve, any small perturbations stay static because both $v_{\text{m}}$ and $v_{\text{dc}}$ vanish. When $|\tau_{\text{e}}|$ slightly exceeds $\tau_{\text{f}}$ (the dark region), the heterogeneities in dislocation density only grow when $\rho_{\text{tot}}s<1/\sqrt{2}$. However, such instability disappears as the difference between $|\tau_{\text{e}}|$ and $\tau_{\text{f}}$ becomes apparent. \label{fig_phase_diagram}}
\end{figure}

Based on the above numerical results, the stability conditions of the continuum model can be summarised schematically in a phase diagram shown in Fig.~\ref{fig_phase_diagram}.
When $|\tau_{\text{e}}|$ falls below $\tau_{\text{f}}$ which is given by the dashed curve in Fig.~\ref{fig_phase_diagram}, any small perturbations stay static because both $v_{\text{m}}$ and $v_{\text{dc}}$ vanish. When $|\tau_{\text{e}}|$ slightly exceeds $\tau_{\text{f}}$, density heterogeneities only grow for $\rho_{\text{tot}}s<1/\sqrt{2}$ (the dark region in Fig.~\ref{fig_phase_diagram}). However, such instability disappears as the difference between $|\tau_{\text{e}}|$ and $\tau_{\text{f}}$ becomes apparent.

\subsection{Long-time behaviour of the continuum model}
\begin{figure}[!ht]
  \centering
  \subfigure[continuum]{\includegraphics[width = .4\textwidth]{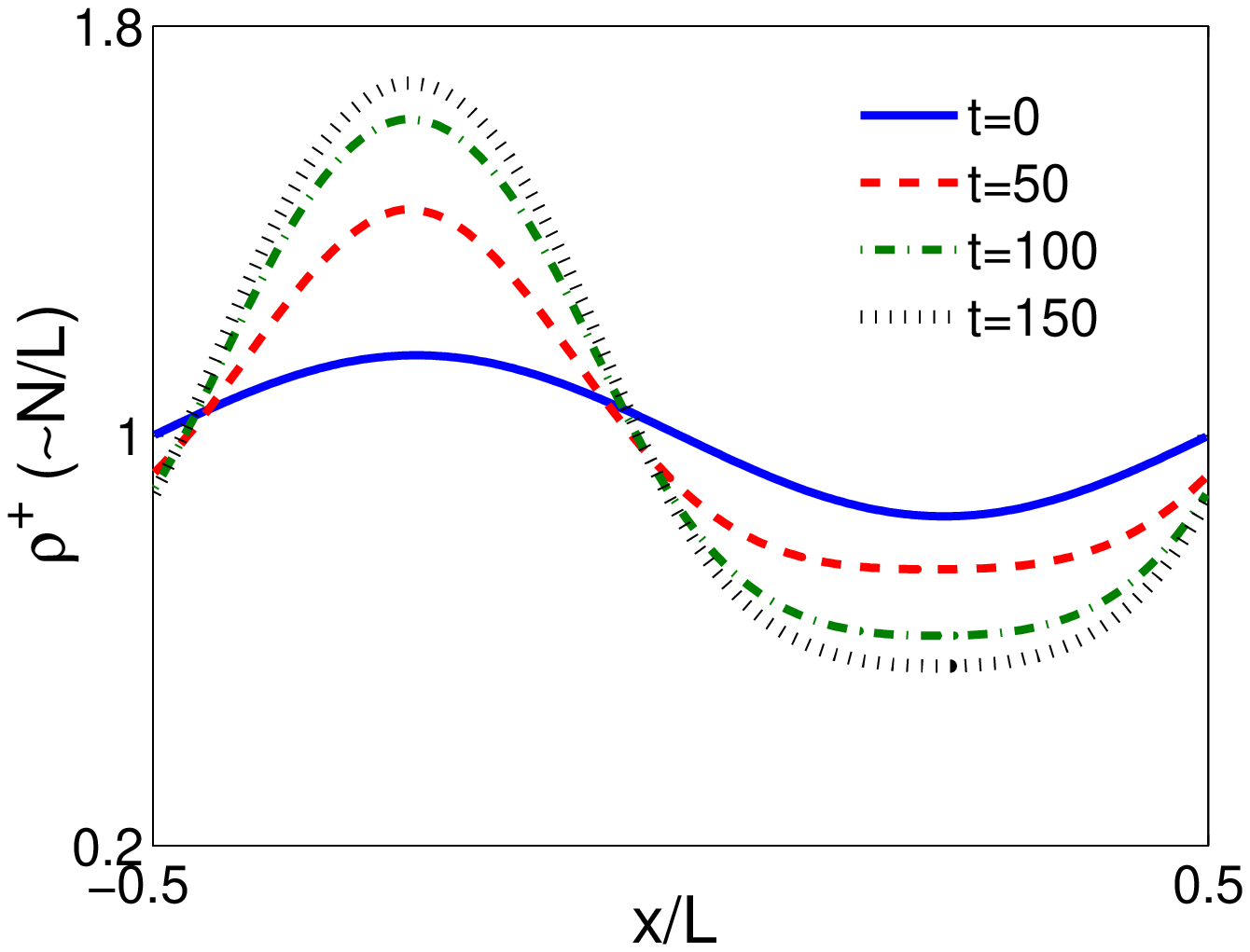}}
  \subfigure[DDD]{\includegraphics[width = .4\textwidth]{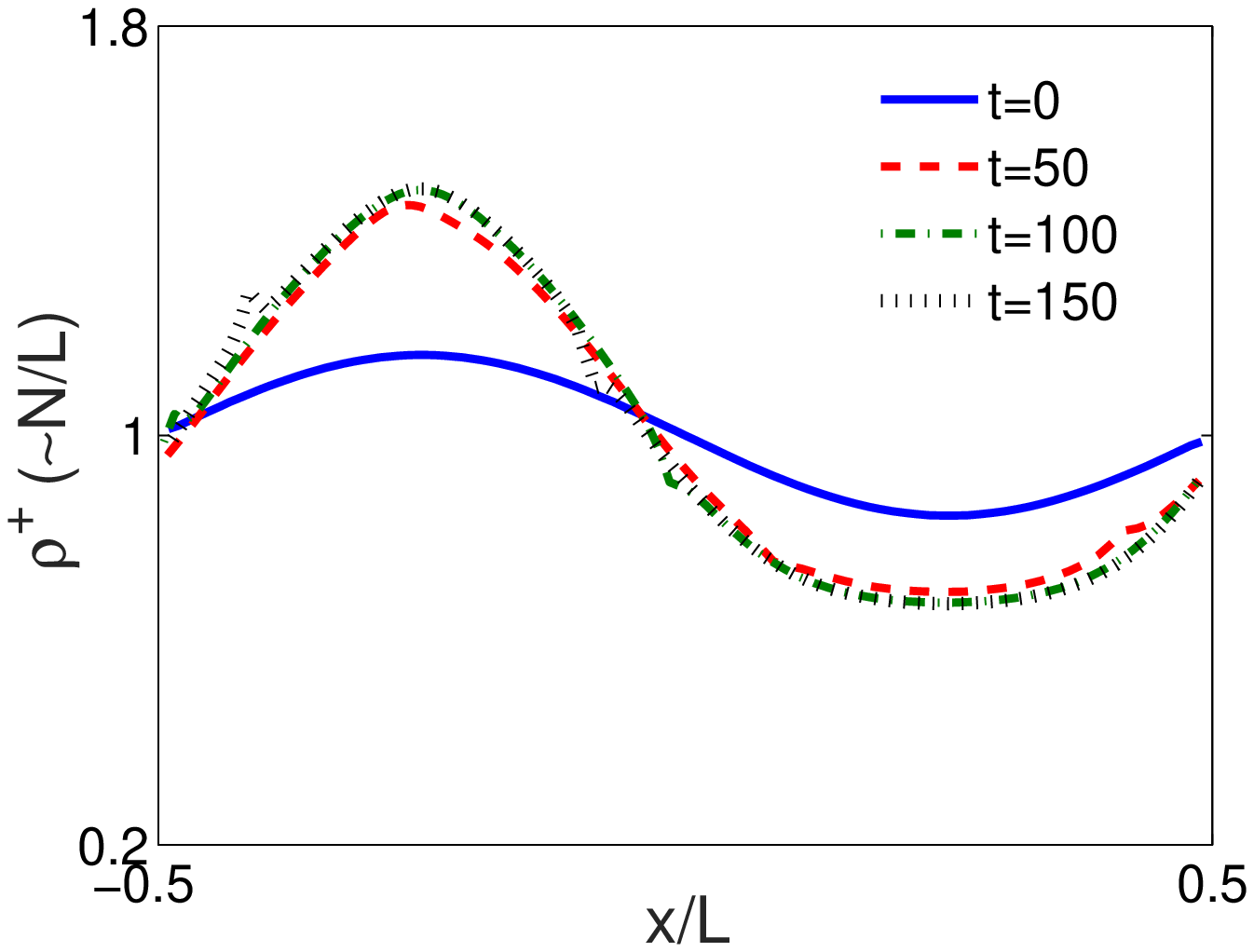}}
  \caption{Long time behaviour of the dislocation system: the density inhomogeneities stop growing at some point. The times shown in the figures are of unit $(1-\nu)L^2/(m_{\text{g}}\mu b^2N)$, $\tau_{\text{e}}=0.24\mu bN/((1-\nu)L)$, $N=100$, $s=100b$ and $L=5\times10^4b$. \label{fig_unstable_longtime}}
\end{figure}

\begin{figure}[!ht]
  \centering
  \subfigure[]{\includegraphics[width=.32\textwidth]{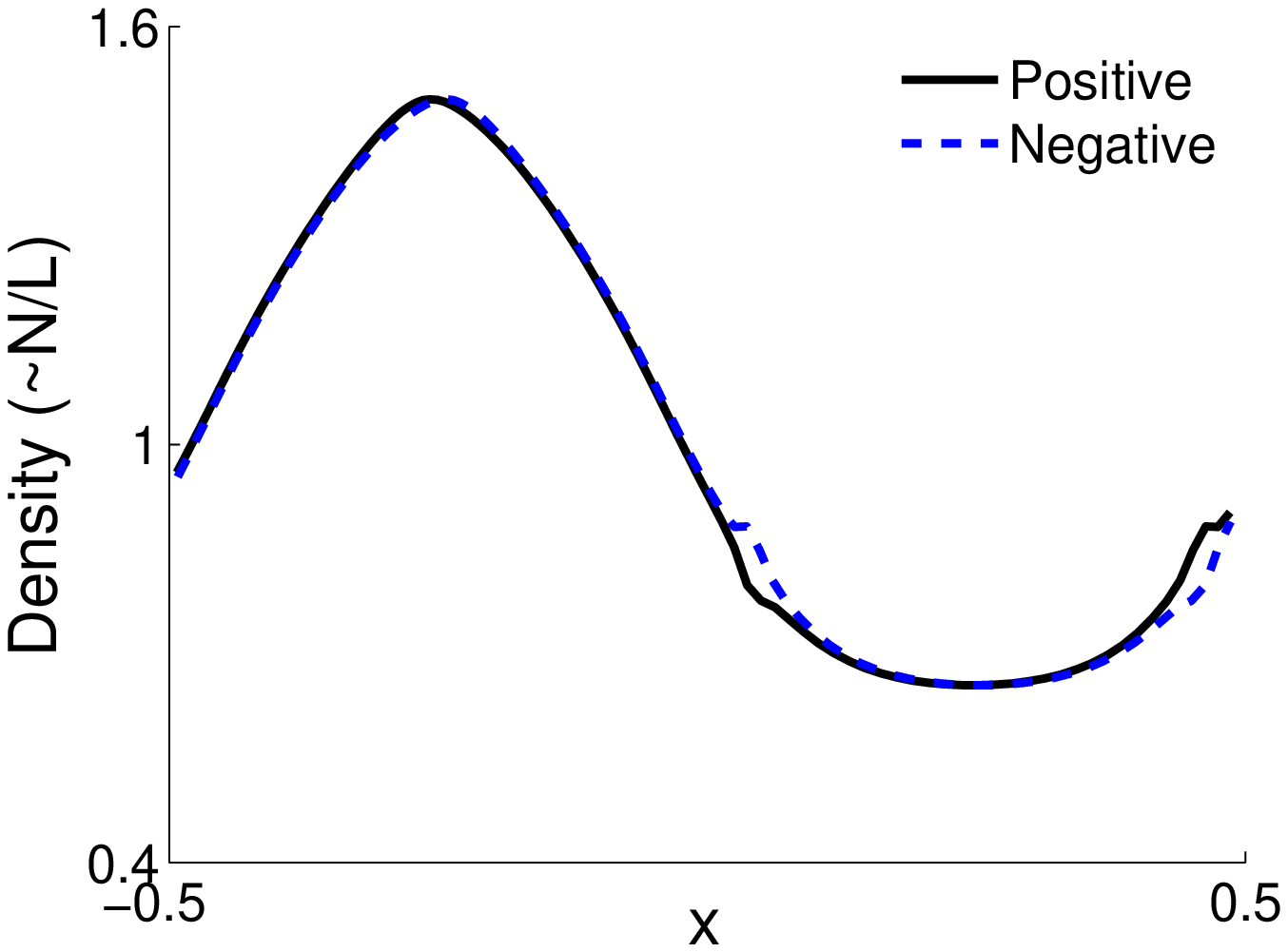}}
  \subfigure[]{\includegraphics[width=.32\textwidth]{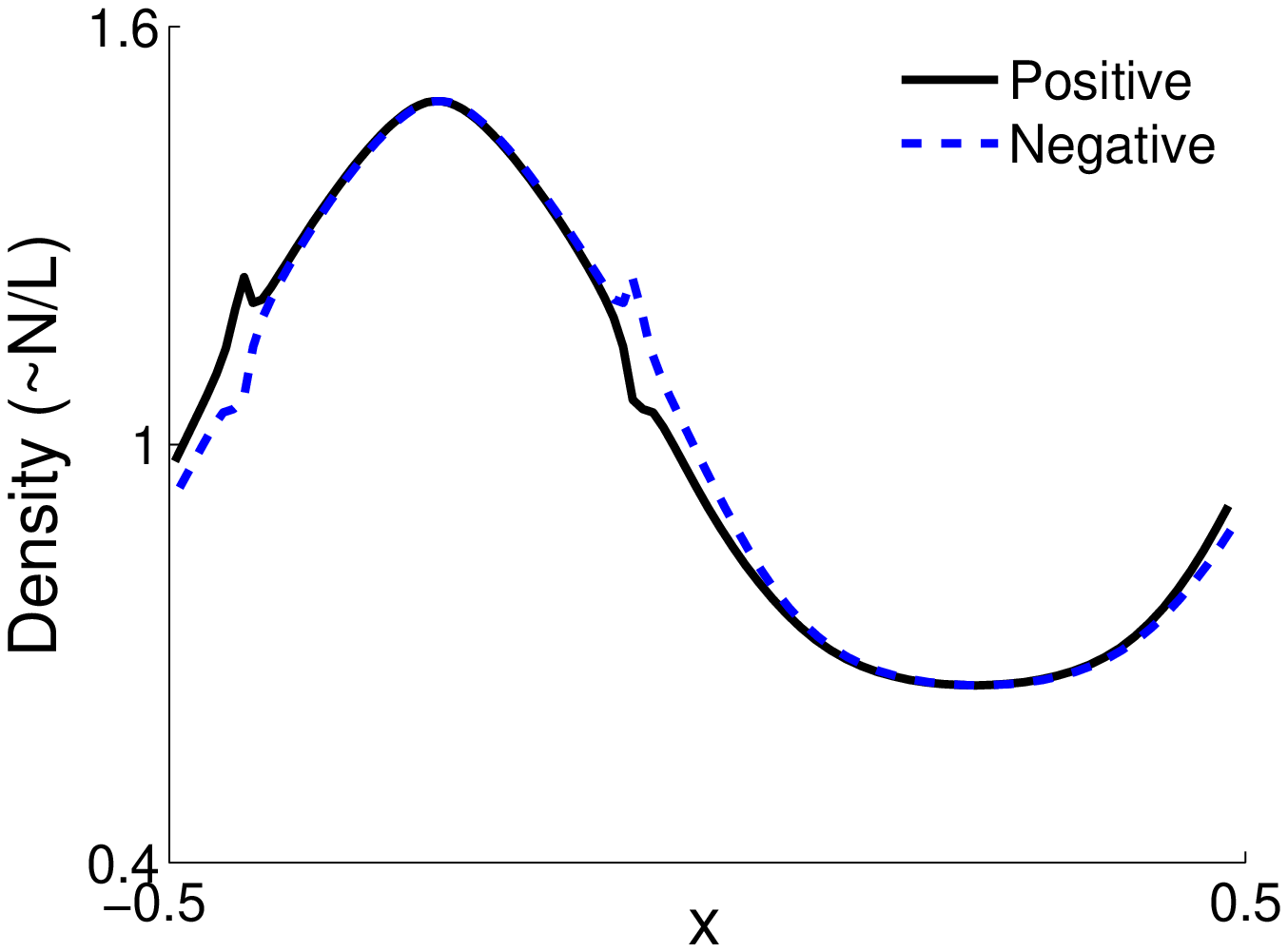}}
  \subfigure[]{\includegraphics[width=.32\textwidth]{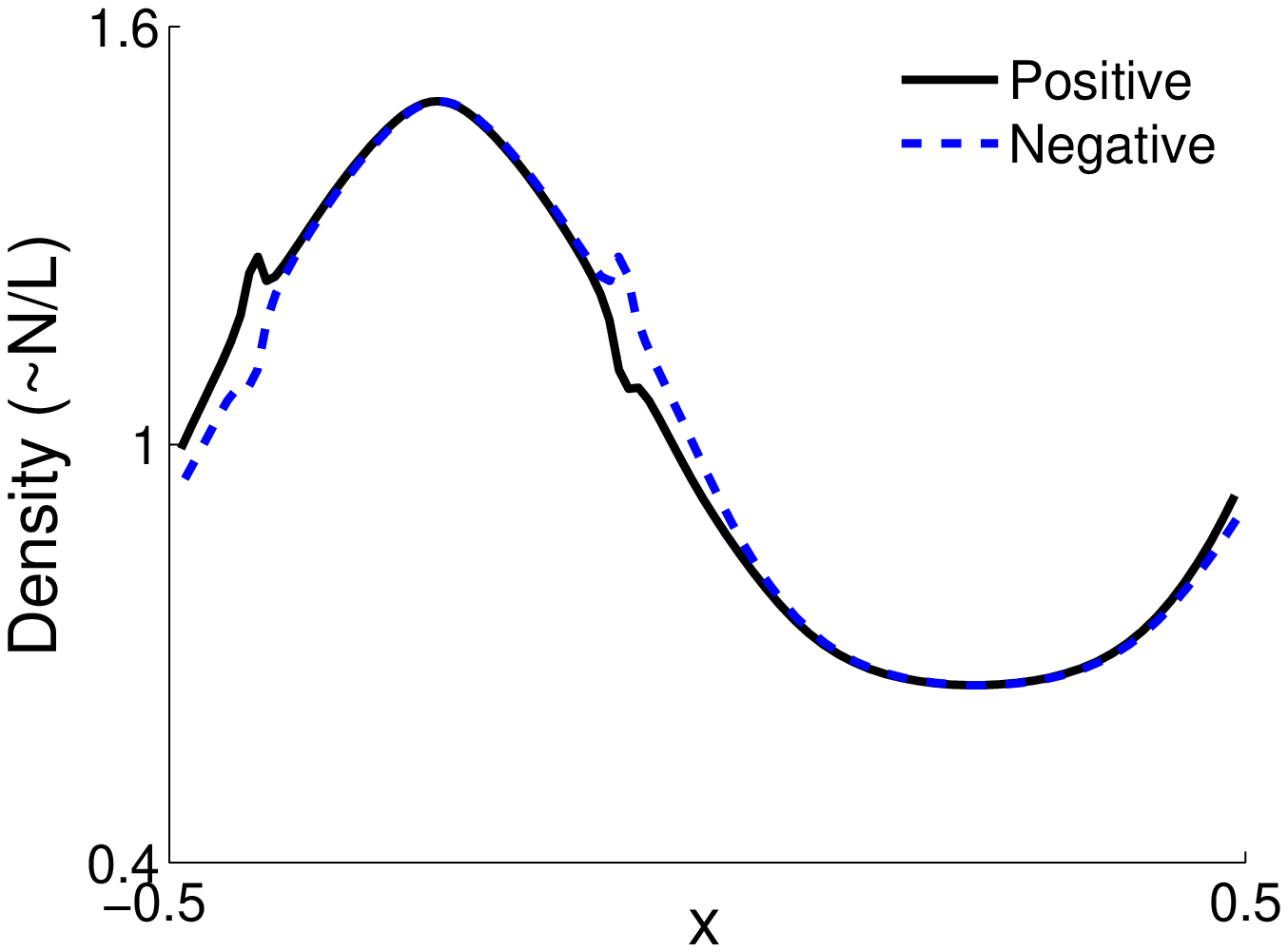}}
  \subfigure[]{\includegraphics[width=.32\textwidth]{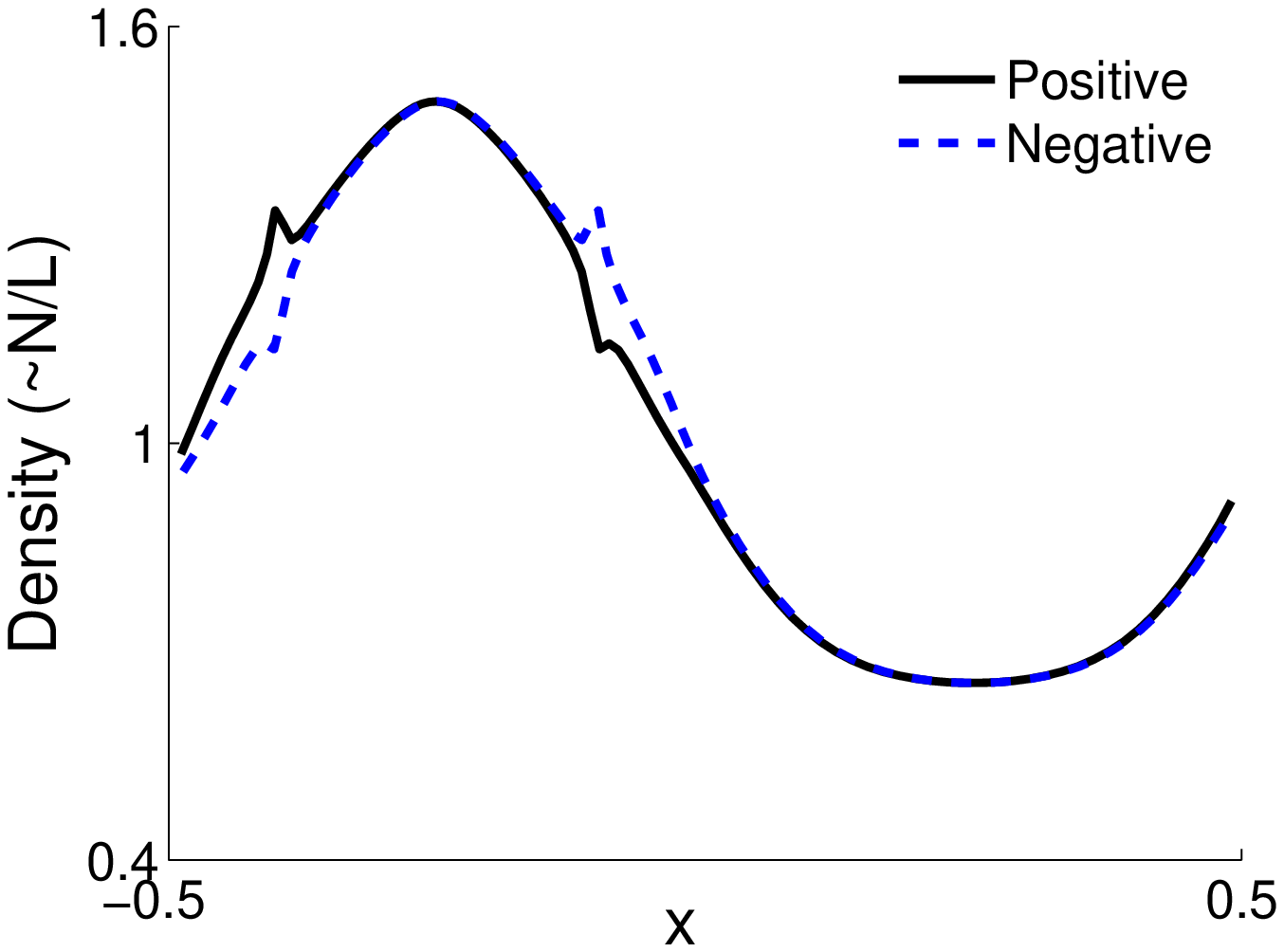}}
  \subfigure[]{\includegraphics[width=.32\textwidth]{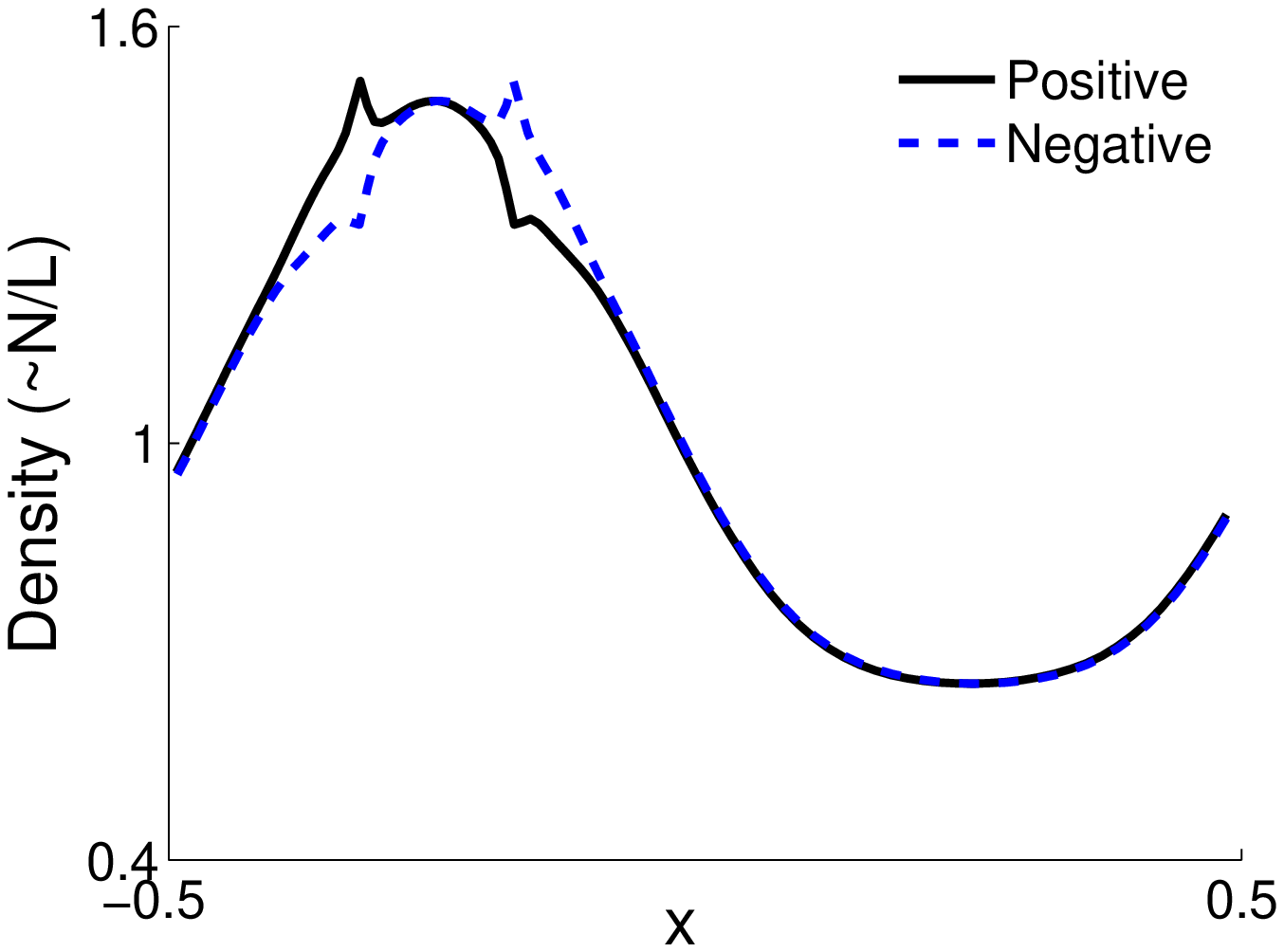}}
  \subfigure[]{\includegraphics[width=.32\textwidth]{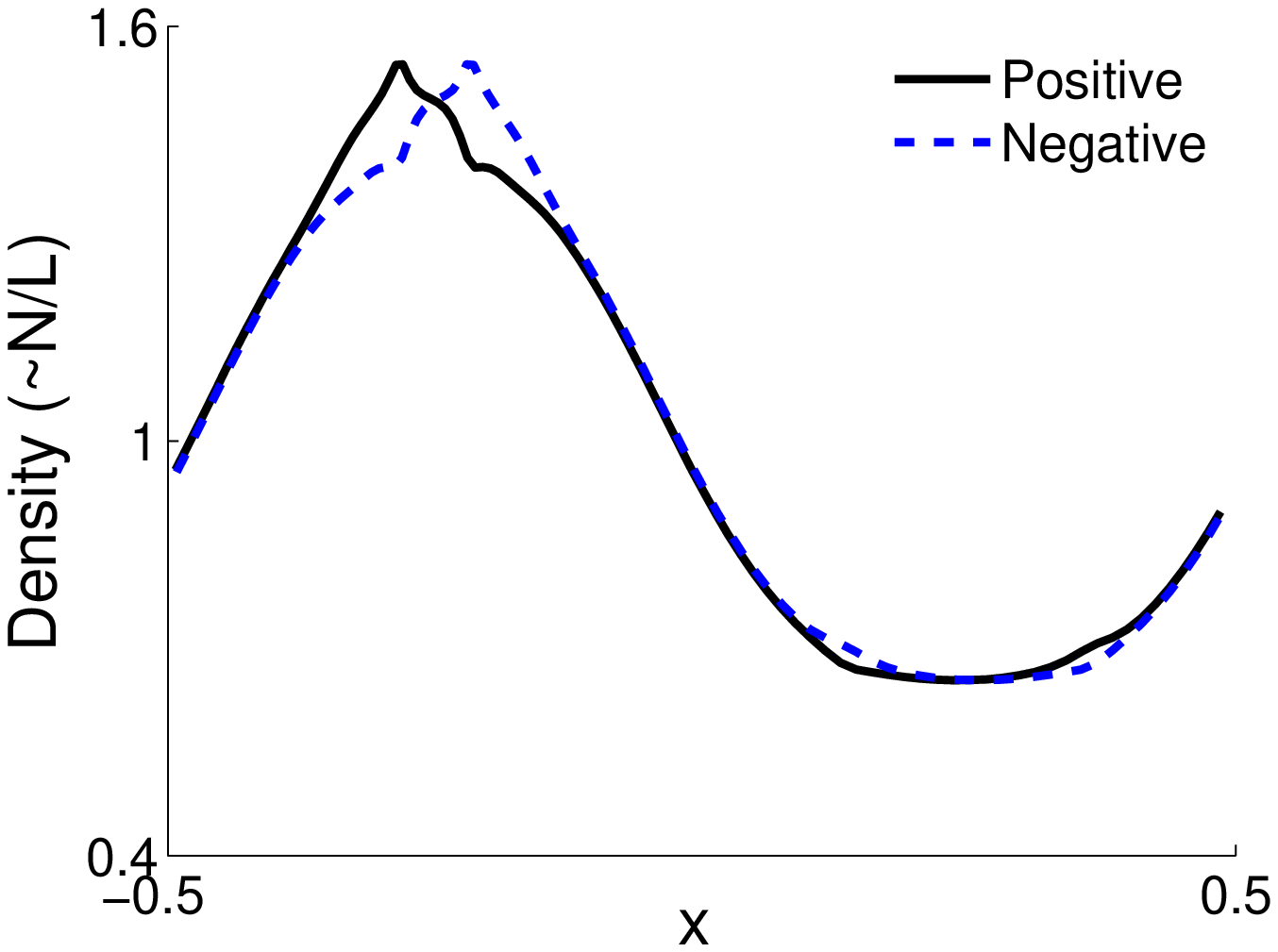}}
  \caption{DDD simulations suggest that the dislocation system eventually reaches a cyclic equilibrium state. \label{fig_ddd_cycle}}
\end{figure}

Now we will show that the derived continuum model is also able to properly capture the long-time behaviour of the underlying discrete dislocation systems. We adopt the same set-up as used for generating Fig.~\ref{fig_instability} ($\tau_{\text{e}}=0.24\mu bN/((1-\nu)L)$, $N=100$, $s=100b$ and $L=5\times10^4b$), and let the evolution time increase 30 times to $150(1-\nu)L^2/(m_{\text{g}}\mu b^2N)$. The long-time behaviour by DDD is shown in Fig.~\ref{fig_unstable_longtime}(b),
which indicates the heterogeneities in dislocation density cease to grow at some point. By taking several snap shots of the discrete system after $t=100(1-\nu)L^2/(m_{\text{g}}\mu b^2N)$, we find that the dislocation system reaches a cyclic equilibrium state as summarised in Fig.~\eqref{fig_ddd_cycle}.
Suppose each cycle starts with an almost GND-free state, then the weaker part ($x\in[0,L/2]$) becomes the source for GNDs (Fig.~\ref{fig_ddd_cycle}(a)). In response to the external stress, GNDs move towards the high-density region from both sides (Fig.~\ref{fig_ddd_cycle}(b)-(d)). When the spiky-looking GND densities are about to join at the peak (Fig.~\ref{fig_ddd_cycle}(e)-(f)), they generate a relatively strong back stress at the peak, which helps break up the dislocation dipoles in the high-density region. As a result, the system evolves back to the almost GND-free state (Fig.~\ref{fig_ddd_cycle}(a)). The dislocation behaviour under the same condition is also studied by the continuum model, and the result shown in Fig.~\ref{fig_unstable_longtime}(a) indicates that the density heterogeneities cease to grow after reaching a statically equilibrium state. The peak value is slightly higher than that in DDD results. Note that since the continuum model captures the average effect of the DDD model, the relatively strong back stress at the peak induced by spiky-looking GND densities in DDD simulations is accommodated by the high-order term in the continuum model.

\section{A general strategy for upscaling dislocation dynamics involving SLDSs\label{Sec_discussion}}
In this section, we analyse the findings in the present work and in other relevant works \citep{dipole_SIAP2016, Zhu_Scripta2016}, so as to summarise a general strategy for upscaling the dynamics of dislocation systems involving SLDSs, as outlined in the end of this section.

First, in order to properly formulate the multi-scale interaction of dislocations, we treat a given dislocation system as a dislocation continua, at each spatial point of which a subcell is attached (see Fig.~\ref{fig_stress_decomposition}). Within the subcell, the positions of discrete dislocations are resolved. The size of the subcell is small when viewed at the continuum level, because the formation of SLDSs is mainly induced by the short-range dislocation correlations. With the introduction of such subcells, the stress field calculated in DDD models can be asymptotically reproduced by the sum of a mean-field stress and a locally discrete stress which is periodic over the subcell. How to choose such a subcell is elucidated in \S~\ref{Sec_con_approx}, and periodic boundary conditions are imposed to the subcell, such that the matching between the locally discrete stress and the mean-field stress is conducted in a more seamless manner. Note that the stress decomposition in this way has been justified both analytically and numerically in the case of a row of dislocation dipoles \citep{dipole_SIAP2016}. The mean-field stress which is considered as a constant inside the subcell, only depends on the distribution of GNDs and loading conditions, and it can be computed, for example, based on some finite element schemes \citep[e.g.][]{Roy2005,Zhu_continuum3D}.

Second, although new variables resolving the positions of discrete dislocations are introduced by the above-mentioned stress decomposition, the dislocation dynamics can still be described by the evolution equations for dislocation densities. This is due to the fact that the subcell is small viewed at a continuum level. Driven by the (singularly) strong short-range dislocation interactions, the self-adjustment of dislocations within a (very small) subcell should take place much more quickly time scale than the evolution of dislocation densities. Hence the evolution of the SLDSs can be treated as a quasi-steady process on the time scale over which dislocation densities evolve normally. For example in this article, in order to determine the decoupling velocity of dislocation pairs $v_{\text{dc}}$ which is defined by Eq.~\eqref{v_sl_def_ddd}, the positions of the discrete dislocations inside the subcell should be tracked. However, since the DDD inside the (periodic) subcell quickly evolves to a (cyclicly) steady state (see Fig.~\ref{fig_3particles}), $v_{\text{dc}}$ can be determined as the (temporal and spatial) average of the velocities of discrete dislocations, which are in cyclicly steady states. The velocity field $v_{\text{dc}}$ is a function of only mean-field quantities, such as the dislocation densities and the local mean-field stress. The influence of the SLDSs is thus formulated as frictional stresses at the continuum level (e.g. see Eq.~\eqref{v_sl_tau_general}). Here one key finding which dramatically simplifies the average process is that, the intra-subcell dislocation interaction is formulated of a Newton's third law type (ref. Eq.~\eqref{tau_per_odd}). As a result, the average dislocation velocity of one species is independent of the intra-species interactions of dislocations inside the subcell (ref. Eq.~\eqref{eqn_semi_continuum_sum_positive}). This is also why the average velocity of all dislocations inside a subcell $v_{\text{m}}$ is independent of the short-range dislocation interactions. One can expect that in the case where there are more parallel slip planes, the corresponding $v_{\text{m}}$ should also only depend on mean-field quantities. In fact, this average-and-cancel technique has also been used for studying the SLDSs formed in multi-slip systems \citep{Zhu_Scripta2016}. Note that in this article, a linear relation between $v_{\text{dc}}$ and $\tau_{\text{m}}$ is assumed for simplicity.

Third, one way to express the flow stress of an SLDS is to study the bifurcation appearing in the long-time behaviour of the DDD inside the subcell. For example as shown in Fig.~\ref{fig_3particles}, the dislocations of opposite sign inside a subcell either drift together at a constant speed or get decoupled intermittently after initial self-adjustment. When the dislocations are mutually-locked and drift together, the velocity of each dislocation inside the subcell should equal the migration velocity of dislocation ensembles, i.e. $V_i^+=V_i^-=v_{\text{m}}$ in Eqs.~\eqref{eqn_DDD_positive} and \eqref{eqn_ddd_negative}. Mathematically, Eqs.~\eqref{eqn_DDD_positive} and \eqref{eqn_ddd_negative} (with $V_i^{\pm}=v_{\text{m}}$) can be considered as an algebraic equation system for the positions of discrete dislocations $\{x_i^{\pm}\}_{i=1}^{N^{\pm}}$ (inside the subcell). The flow stress is thus the maximum mean-field stress under which the solutions (given by $\{x_i^{\pm}\}_{i=1}^{N^{\pm}}$) to Eqs.~\eqref{eqn_DDD_positive} and \eqref{eqn_ddd_negative} (with $V_i^{\pm}=v_{\text{m}}$) still exist. Analysing the solvability conditions of the algebraic equation system gives rise to Eq.~\eqref{DDD_fric_max}, which defines the flow stress $\tau_{\text{f}}$ in this article. Note that the determination of $\tau_{\text{f}}$ in Eq.~\eqref{DDD_fric_max} is associated with an optimisation process over all $\{x_i^{\pm}\}_{i=1}^{N^{\pm}}$. Hence the flow stress is only dependent on mean-field quantities. This principle by relating the flow stress to the solvability conditions of DDD equations, has also been implemented to derive the flow stress formulae of SLDSs in multi-slip systems \citep{Zhu_Scripta2016}.

Based on the discussion above, a general strategy for incorporating SLDSs into continuum frameworks of dislocations is summarised in the following three principles:
\begin{enumerate}
  \item[i.] A given dislocation system can be considered as a dislocation continuum, at each spatial point of which a subcell is embedded, and the stress calculated in DDD models can be asymptotically decomposed to a sum of a mean-field stress and a locally periodic discrete stress.
  \item[ii.] The dislocation dynamics can be formulated by a set of evolution equations for dislocation densities, while the influence of SLDSs is taken into account quasi-steadily as frictional stresses.
  \item[iii.] The flow stress associated with an SLDS can be determined by checking the solvability conditions of the locally DDD equations, which correspond to the case where all dislocations inside the subcell are locked together.
\end{enumerate}

Under the above-mentioned principles, an effective linkage between the materials macroscopic properties such as the yield stress and the underlying SLDSs can be established, where the mean-field stress field acts as the media.

\section{Summary\label{Sec_summary}}
To summarise, a continuum model is derived to describe the collective behaviour of arbitrary combinations of straight edge dislocations on two parallel slip planes. In order to facilitate the discrete-to-continuum transition, two continuum velocity fields are introduced: the average dislocation migration velocity $v_{\text{m}}$ and the decoupling velocity of dislocation pairs $v_{\text{dc}}$. By analysing the underlying discrete dynamics, $v_{\text{m}}$ is found independent of the short-range elastic interactions of dislocations, while $v_{\text{dc}}$ bifurcates controlled by a critical stress $\tau_{\text{f}}$ which is termed as the effective flow stress here. Compared to the widely used Taylor-type flow stress formulae, the expression of $\tau_{\text{f}}$ derived in this article takes into account the anisotropic behaviour of dislocations, and automatically admits a minimum average inter-dislocation distance within the slip planes. Two well formulated cases are found special cases of the derived model. With the continuum model, a regime under which the density heterogeneities of dislocations grow is identified, and the result is validated through a comparison with DDD simulations. The derived continuum model is also shown capable of effectively capturing the long-time behaviour of the underlying discrete systems.

The presented work represents a first and also reasonably successful attempt of implementing the general strategy outlined in \S~\ref{Sec_discussion} to the continuum formulation of the dynamical processes involving the formation, migration and dissociation of SLDSs. For systematically building plasticity theory of dislocation continua, more complicated situations where dislocations are curved and there are multiple slip planes must be properly formulated, which will be explored in the future work.

\section*{Acknowledgement}
This work is partially supported by the Hong Kong Research Grants Council General
Research Fund 606313.

\bibliography{mybib}

\begin{thebibliography}{44}
\providecommand{\natexlab}[1]{#1}
\providecommand{\url}[1]{\texttt{#1}}
\expandafter\ifx\csname urlstyle\endcsname\relax
  \providecommand{\doi}[1]{doi: #1}\else
  \providecommand{\doi}{doi: \begingroup \urlstyle{rm}\Url}\fi

\bibitem[Acharya(2001)]{Acharya2001}
A.~Acharya.
\newblock A model of crystal plasticity based on the theory of continuously
  distributed dislocations.
\newblock \emph{J. Mech. Phys. Solids}, 49\penalty0 (4):\penalty0 761--784,
  2001.

\bibitem[Akarapu et~al.(2010)Akarapu, Zbib, and Bahr]{Zbib2010}
S.~Akarapu, H.~M. Zbib, and D.~F. Bahr.
\newblock Analysis of heterogeneous deformation and dislocation dynamics in
  single crystal micropillars under compression.
\newblock \emph{Int. J. Plast.}, 26\penalty0 (2):\penalty0 239--257, 2010.

\bibitem[Arsenlis and Parks(2002)]{Arsenlis2002}
A.~Arsenlis and D.~M. Parks.
\newblock Modeling the evolution of crystallographic dislocation density in
  crystal plasticity.
\newblock \emph{J. Mech. Phys. Solids}, 50\penalty0 (9):\penalty0 1979--2009,
  2002.

\bibitem[Bulatov et~al.(2006)Bulatov, Hsiung, Tang, Arsenlis, Bartelt, Cai,
  Florando, Hiratani, Rhee, Hommes, Pierce, and de~la
  Rubia]{Bulatov_Nature2006}
V.~V. Bulatov, L.~L. Hsiung, M.~Tang, A.~Arsenlis, M.~C. Bartelt, W.~Cai, J.~N.
  Florando, M.~Hiratani, M.~Rhee, G.~Hommes, T.~G. Pierce, and T.~D. de~la
  Rubia.
\newblock Dislocation multi-junctions and strain hardening.
\newblock \emph{Nature}, 440\penalty0 (7088):\penalty0 1174--1178, 2006.

\bibitem[Chapman et~al.(2016)Chapman, Xiang, and Zhu]{dipole_SIAP2016}
S.~J. Chapman, Y.~Xiang, and Y.~C. Zhu.
\newblock Homogenization of a row of dislocation dipoles from discrete
  dislocation dynamics.
\newblock \emph{Siam J. Appl. Math.}, 76\penalty0 (2):\penalty0 750--775, 2016.

\bibitem[Duong et~al.(2015)Duong, Lamacz, Peletier, and Sharma]{Peletier2015}
M.~H. Duong, A.~Lamacz, M.~A. Peletier, and U.~Sharma.
\newblock Variational approach to coarse-graining of generalized gradient
  flows.
\newblock 2015.
\newblock URL \url{arxiv.org/abs/1507.06322}.

\bibitem[El-Awady(2015)]{ElAlwardy_NatCom2015}
J.~A. El-Awady.
\newblock Unravelling the physics of size-dependent dislocation-mediated
  plasticity.
\newblock \emph{Nat. Commun.}, 6, 2015.

\bibitem[El-Azab(2000)]{ElAzab2000}
A.~El-Azab.
\newblock Statistical mechanics treatment of the evolution of dislocation
  distributions in single crystals.
\newblock \emph{Phys. Rev. B}, 61\penalty0 (18):\penalty0 11956--11966, 2000.

\bibitem[Franciosi and Zaoui(1982)]{Franciosi1982}
P.~Franciosi and A.~Zaoui.
\newblock Multislip in f.c.c. crystals a theoretical approach compared with
  experimental data.
\newblock \emph{Acta Metall.}, 30\penalty0 (8):\penalty0 1627--1637, 1982.

\bibitem[Geers et~al.(2013)Geers, Peerlings, Peletier, and Scardia]{Geers2013}
M.~G.~D. Geers, R.~H.~J. Peerlings, M.~A. Peletier, and L.~Scardia.
\newblock Asymptotic behaviour of a pile-up of infinite walls of edge
  dislocations.
\newblock \emph{Arch. Ration. Mech. Anal.}, 209\penalty0 (2):\penalty0
  495--539, 2013.

\bibitem[Groma(1997)]{Groma1997}
I.~Groma.
\newblock Link between the microscopic and mesoscopic length-scale description
  of the collective behavior of dislocations.
\newblock \emph{Phys. Rev. B}, 56\penalty0 (10):\penalty0 5807--5813, 1997.

\bibitem[Groma and Balogh(1999)]{Groma1999}
I.~Groma and P.~Balogh.
\newblock Investigation of dislocation pattern formation in a two-dimensional
  self-consistent field approximation.
\newblock \emph{Acta Mater.}, 47\penalty0 (13):\penalty0 3647--3654, 1999.

\bibitem[Groma et~al.(2003)Groma, Csikor, and Zaiser]{Groma2003}
I.~Groma, F.~F. Csikor, and M.~Zaiser.
\newblock Spatial correlations and higher-order gradient terms in a continuum
  description of dislocation dynamics.
\newblock \emph{Acta Mater.}, 51\penalty0 (5):\penalty0 1271--1281, 2003.

\bibitem[Groma et~al.(2006)Groma, Gy\"{o}rgyi, and Kocsis]{Groma2006}
I.~Groma, G.~Gy\"{o}rgyi, and B.~Kocsis.
\newblock Debye screening of dislocations.
\newblock \emph{Phys. Rev. Lett.}, 96\penalty0 (16), 2006.

\bibitem[Groma et~al.(2010)Groma, Gy\"{o}rgyi, and Isp\'{a}novity]{Groma2010}
I.~Groma, G.~Gy\"{o}rgyi, and P.~D. Isp\'{a}novity.
\newblock Variational approach in dislocation theory.
\newblock \emph{Phil. Mag.}, 90\penalty0 (27-28):\penalty0 3679--3695, 2010.

\bibitem[Groma et~al.(2016)Groma, Zaiser, and Isp\'{a}novity]{Groma2016}
I.~Groma, M.~Zaiser, and P.~D. Isp\'{a}novity.
\newblock Dislocation patterning in a two-dimensional continuum theory of
  dislocations.
\newblock \emph{Phys. Rev. B}, 93\penalty0 (21):\penalty0 214110, 2016.

\bibitem[Head(1972)]{Head1972c}
A.~K. Head.
\newblock Dislocation group dynamics iii. similarity solutions of continuum
  approximation.
\newblock \emph{Phil. Mag.}, 26\penalty0 (1):\penalty0 65--72, 1972.

\bibitem[Hochrainer et~al.(2007)Hochrainer, Zaiser, and
  Gumbsch]{Hochrainer2007}
T.~Hochrainer, M.~Zaiser, and P.~Gumbsch.
\newblock A three-dimensional continuum theory of dislocation systems:
  kinematics and mean-field formulation.
\newblock \emph{Phil. Mag.}, 87\penalty0 (8-9):\penalty0 1261--1282, 2007.

\bibitem[Hochrainer et~al.(2014)Hochrainer, Sandfeld, Zaiser, and
  Gumbsch]{Hochrainer2014}
T.~Hochrainer, S.~Sandfeld, M.~Zaiser, and P.~Gumbsch.
\newblock Continuum dislocation dynamics: Towards a physical theory of crystal
  plasticity.
\newblock \emph{J. Mech. Phys. Solids}, 63:\penalty0 167--178, 2014.

\bibitem[Kooiman et~al.(2014)Kooiman, H\"{u}tter, and Geers]{Kooiman2014}
M.~Kooiman, M.~H\"{u}tter, and M.~G.~D. Geers.
\newblock Collective behaviour of dislocations in a finite medium.
\newblock \emph{J. Stat. Mech.}, 2014\penalty0 (4):\penalty0 P04028, 2014.

\bibitem[Kooiman et~al.(2015)Kooiman, H\"{u}tter, and Geers]{Geers_JMPS2015}
M.~Kooiman, M.~H\"{u}tter, and M.~G.~D. Geers.
\newblock Microscopically derived free energy of dislocations.
\newblock \emph{J. Mech. Phys. Solids}, 78:\penalty0 186--209, 2015.

\bibitem[Kroener(1958)]{Kroener1958}
E.~Kroener.
\newblock \emph{Kontinnuumstheorie der Bersetzungen und Eigenspannungen}.
\newblock Springer-Verlag, Berlin, 1958.

\bibitem[Leung and Ngan(2016)]{Ngan_JMPS2016}
H.~S. Leung and A.~H.~W. Ngan.
\newblock Dislocation-density function dynamics ¨c an all-dislocation,
  full-dynamics approach for modeling intensive dislocation structures.
\newblock \emph{J. Mech. Phys. Solids}, 2016.

\bibitem[Leung et~al.(2015)Leung, Leung, Cheng, and Ngan]{Ngan2015}
H.~S. Leung, P.~S.~S. Leung, B.~Cheng, and A.~H.~W. Ngan.
\newblock A new dislocation-density-function dynamics scheme for computational
  crystal plasticity by explicit consideration of dislocation elastic
  interactions.
\newblock \emph{Int. J. Plast.}, 67:\penalty0 1--25, 2015.

\bibitem[Madec et~al.(2002{\natexlab{a}})Madec, Devincre, and Kubin]{Madec2002}
R.~Madec, B.~Devincre, and L.~P. Kubin.
\newblock Simulation of dislocation patterns in multislip.
\newblock \emph{Scripta Mater.}, 47\penalty0 (10):\penalty0 689--695,
  2002{\natexlab{a}}.

\bibitem[Madec et~al.(2002{\natexlab{b}})Madec, Devincre, and
  Kubin]{Madec_PRL2002}
R.~Madec, B.~Devincre, and L.~P. Kubin.
\newblock From dislocation junctions to forest hardening.
\newblock \emph{Phys. Rev. Lett.}, 89\penalty0 (25), 2002{\natexlab{b}}.

\bibitem[Mielke(2011)]{Mielke2011}
A.~Mielke.
\newblock Formulation of thermoelastic dissipative material behavior using
  generic.
\newblock \emph{Contin. Mech. Thermodyn.}, 23\penalty0 (3):\penalty0 233--256,
  2011.

\bibitem[Mielke and Ortiz(2008)]{Mielke2008}
A.~Mielke and M.~Ortiz.
\newblock A class of minimum principles for characterizing the trajectories and
  the relaxation of dissipative systems.
\newblock \emph{ESAIM: COCV}, 14\penalty0 (3):\penalty0 494--516, 2008.

\bibitem[Monavari et~al.(2016)Monavari, Sandfeld, and Zaiser]{Monavari2016}
M.~Monavari, S.~Sandfeld, and M.~Zaiser.
\newblock Continuum representation of systems of dislocation lines: A general
  method for deriving closed-form evolution equations.
\newblock \emph{J. Mech. Phys. Solids}, 2016.

\bibitem[Mughrabi(1976)]{Mughrabi1976}
H.~Mughrabi.
\newblock Observation of pinned dislocation arrangements by transmissions
  electron microscopy (tem).
\newblock \emph{J. Microsc. Spectrosc. Electron}, 1:\penalty0 571 -- 584, 1976.

\bibitem[Nye(1953)]{Nye1953}
J.~F. Nye.
\newblock Some geometrical relations in dislocated crystals.
\newblock \emph{Acta Metall.}, 1:\penalty0 153 -- 162, 1953.

\bibitem[Roy and Acharya(2005)]{Roy2005}
A.~Roy and A.~Acharya.
\newblock Finite element approximation of field dislocation mechanics.
\newblock \emph{J. Mech. Phys. Solids}, 53\penalty0 (1):\penalty0 143--170,
  2005.

\bibitem[Sandfeld and Zaiser(2015)]{Sandfeld_pattern2015}
S.~Sandfeld and M.~Zaiser.
\newblock Pattern formation in a minimal model of continuum dislocation
  plasticity.
\newblock \emph{Modeling Simul. Mater. Sci. Eng.}, 23\penalty0 (6):\penalty0
  065005, 2015.

\bibitem[Schmitt et~al.(2015)Schmitt, Gumbsch, and Schulz]{Katrin_JMPS2015}
S.~Schmitt, P.~Gumbsch, and K.~Schulz.
\newblock Internal stresses in a homogenized representation of dislocation
  microstructures.
\newblock \emph{J. Mech. Phys. Solids}, 84:\penalty0 528--544, 2015.

\bibitem[Schulz et~al.(2014)Schulz, Dickel, Schmitt, Sandfeld, Weygand, and
  Gumbsch]{Schulz2014}
K.~Schulz, D.~Dickel, S.~Schmitt, S.~Sandfeld, D.~Weygand, and P.~Gumbsch.
\newblock Analysis of dislocation pile-ups using a dislocation-based continuum
  theory.
\newblock \emph{Modelling Simul. Mater. Sci. Eng.}, 22\penalty0 (2), 2014.

\bibitem[Taylor(1934)]{Taylor1934}
G.~I. Taylor.
\newblock The mechanism of plastic deformation of crystals. part i.
  theoretical.
\newblock \emph{Proc. Roy. Soc.}, 145\penalty0 (855):\penalty0 362--387, 1934.

\bibitem[Valdenaire et~al.(2016)Valdenaire, Le~Bouar, Appolaire, and
  Finel]{Finel2016}
P.~L. Valdenaire, Y.~Le~Bouar, B.~Appolaire, and A.~Finel.
\newblock Density-based crystal plasticity: From the discrete to the continuum.
\newblock \emph{Phys. Rev. B}, 93\penalty0 (21):\penalty0 214111, 2016.

\bibitem[Xiang(2009)]{Xiang2009_JMPS}
Y.~Xiang.
\newblock Continuum approximation of the peach-koehler force on dislocations in
  a slip plane.
\newblock \emph{J. Mech. Phys. Solids}, 57\penalty0 (4):\penalty0 728--743,
  2009.

\bibitem[Zaiser(2013)]{Zaiser2013}
M.~Zaiser.
\newblock The energetics and interactions of random dislocation walls.
\newblock \emph{Phil. Mag. Lett.}, 93\penalty0 (7):\penalty0 387--394, 2013.

\bibitem[Zhang et~al.(2015)Zhang, Acharya, Walkington, and Bielak]{Acharya2015}
X.~Zhang, A.~Acharya, N.~J. Walkington, and J.~Bielak.
\newblock A single theory for some quasi-static, supersonic, atomic, and
  tectonic scale applications of dislocations.
\newblock \emph{J. Mech. Phys. Solids}, 84:\penalty0 145--195, 2015.

\bibitem[Zhu and Chapman(2014)]{Zhu_2Ddipoles2014}
Y.~C. Zhu and S.~J. Chapman.
\newblock A natural transition between equilibrium patterns of dislocation
  dipoles.
\newblock \emph{J. Elast.}, 117\penalty0 (1):\penalty0 51--61, 2014.

\bibitem[Zhu and Xiang(2015)]{Zhu_continuum3D}
Y.~C. Zhu and Y.~Xiang.
\newblock A continuum model for dislocation dynamics in three dimensions using
  the dislocation density potential functions and its application to
  micro-pillars.
\newblock \emph{J. Mech. Phys. Solids}, 84:\penalty0 230--253, 2015.

\bibitem[Zhu et~al.(2014)Zhu, Wang, Zhu, and Xiang]{Zhu2014_IJP}
Y.~C. Zhu, H.~Q. Wang, X.~H. Zhu, and Y.~Xiang.
\newblock A continuum model for dislocation dynamics incorporating frank¨cread
  sources and hall¨cpetch relation in two dimensions.
\newblock \emph{Int. J. Plast.}, 60\penalty0 (0):\penalty0 19--39, 2014.

\bibitem[Zhu et~al.(2016)Zhu, Xiang, and Schulz]{Zhu_Scripta2016}
Y.~C. Zhu, Y.~Xiang, and K.~Schulz.
\newblock The role of dislocation pile-up in flow stress determination and
  strain hardening.
\newblock \emph{Scripta Mater.}, 116:\penalty0 53--56, 2016.

\end{thebibliography}

\end{document}